# Asteroid (101955) Bennu in the Laboratory: Properties of the Sample Collected by OSIRIS-REx

Dante S. Lauretta[1]*, Harold C. Connolly, Jr.[1,2,3]*, Joseph E. Aebersold[4], Conel M. O'D. Alexander[5], Ronald-L. Ballouz[6], Jessica J. Barnes[1], Helena C. Bates[7], Carina A. Bennett[1], Laurinne Blanche[1], Erika H. Blumenfeld[8], Simon J. Clemett[9], George D. Cody[5], Daniella N. DellaGiustina[1], Jason P. Dworkin[10], Scott A. Eckley[11], Dionysis I. Foustoukos[5], Ian A. Franchi[12], Daniel P. Glavin[10], Richard C. Greenwood[12], Pierre Haenecour[1], Victoria E. Hamilton[13], Dolores H. Hill[1], Takahiro Hiroi[14], Kana Ishimaru[1], Fred Jourdan[15], Hannah H. Kaplan[10], Lindsay P. Keller[16], Ashley J. King[7], Piers Koefoed[17], Melissa K. Kontogiannis[1], Loan Le[11], Robert J. Macke[18], Timothy J. McCoy[19], Ralph E. Milliken[14], Jens Najorka[7], Ann N. Nguyen[16], Maurizio Pajola[20], Anjani T. Polit[1], Heather L. Roper[1], Sara S. Russell[7], Andrew J. Ryan[1], Scott A. Sandford[21], Paul F. Schofield[7], Cody D. Schultz[14], Shogo Tachibana[22], Kathie L. Thomas-Keprta[23], Michelle S. Thompson[24], Valerie Tu[11], Filippo Tusberti[20], Kun Wang[17], Thomas J. Zega[1], C. W. V. Wolner[1], and the OSIRIS-REx Sample Analysis Team[25].

[1]Lunar and Planetary Laboratory, University of Arizona, Tucson, AZ, USA.
[2]Department of Geology, Rowan University, Glassboro, NJ, USA.
[3]Department of Earth and Planetary Science, American Museum of Natural History, New York, NY, USA.
[4]JETS (JSC Engineering & Technical Support) at Texas State University, NASA Johnson Space Center, Houston, TX, USA.
[5]Earth and Planets Laboratory, Carnegie Institution for Science, Washington, DC, USA.
[6]The Johns Hopkins University Applied Physics Laboratory, Laurel, MD, USA.
[7]Natural History Museum, London, UK.
[8]LZ Technology, JETS Contract, NASA Johnson Space Center, Houston, TX, USA.
[9]ERC Inc., NASA Johnson Space Center, Houston, TX, USA.
[10]NASA Goddard Space Flight Center, Greenbelt, MD, USA.
[11]Jacobs, NASA Johnson Space Center, Houston, TX, USA.
[12]School of Physical Sciences, Open University, Milton Keynes, UK.
[13]Southwest Research Institute, Boulder, CO, USA.
[14]Department of Earth, Environmental, and Planetary Sciences, Brown University, Providence, RI, USA.
[15]Space Science and Technology Centre, Curtin University, Perth, Australia.
[16]ARES, NASA Johnson Space Center, Houston, TX, USA.
[17]McDonnell Center for the Space Sciences, Department of Earth, Environmental, & Planetary Sciences, Washington University, St. Louis, MO, USA.
[18]Vatican Observatory, Vatican City State.
[19]National Museum of Natural History, Smithsonian Institution, Washington, DC, USA.
[20] INAF, Astronomical Observatory of Padova, Padova, Italy.
[21]NASA Ames Research Center, Moffett Field, CA, USA.

[22]UTokyo Organization for Planetary and Space Science, The University of Tokyo, Tokyo, Japan.
[23]Barrios Technology/Jacobs, NASA Johnson Space Center, Houston, TX, USA.
[24]Department of Earth, Atmospheric and Planetary Sciences, Purdue University, West Lafayette, IN, USA.
[25]See Appendix.

*These authors contributed equally to this work.

Submitted to Meteoritics and Planetary Science
April 18, 2024

## Table of Contents

## Abstract

On 24 September 2023, NASA's OSIRIS-REx mission dropped a capsule to Earth containing ~120 g of pristine carbonaceous regolith from Bennu. We describe the delivery and initial allocation of this asteroid sample and introduce its bulk physical, chemical, and mineralogical properties from early analyses. The regolith is very dark overall, with higher-reflectance inclusions and particles interspersed. Particle sizes range from sub-micron dust to a stone ~3.5 cm long. Millimeter-scale and larger stones typically have hummocky or angular morphologies. A subset of the stones appears mottled by brighter material that occurs as veins and crusts. Hummocky stones have the lowest densities and mottled stones have the highest. Remote sensing of Bennu's surface detected hydrated phyllosilicates, magnetite, organic compounds, carbonates, and scarce anhydrous silicates, all of which the sample confirms. We also find sulfides, presolar grains, and, less expectedly, Na-rich phosphates, as well as other trace phases. The sample's composition and mineralogy indicate substantial aqueous alteration and resemble those of Ryugu and the most chemically primitive, low-petrologic-type carbonaceous chondrites. Nevertheless, we find distinct hydrogen, nitrogen, and oxygen isotopic compositions, and some of the material we analyzed is enriched in fluid-mobile elements. Our findings underscore the value of sample return — especially for low-density material that may not readily survive atmospheric entry — and lay the groundwork for more comprehensive analyses.

## Introduction

The OSIRIS-REx (Origins, Spectral Interpretation, Resource Identification, and Security–Regolith Explorer) mission returned regolith samples from near-Earth asteroid (101955) Bennu after a seven-year round-trip journey (Lauretta et al., 2017, 2021, 2023a). Bennu, a small (~500 m diameter) B-type asteroid (Figure 1A), was selected as the mission target in part because telescopic observations indicated a primitive, carbonaceous composition and water-bearing minerals (Clark et al., 2011; Lauretta et al., 2015, 2017). These findings suggested that Bennu could contain prebiotic compounds relevant to the origin of life, as well as essential ingredients such as water that could contribute to habitability on Earth and other celestial bodies. Sample return would offer the opportunity to study such primitive material in pristine condition, without the terrestrial contamination that affects meteorites. Understanding Bennu's physical properties was also a motivation owing to its non-zero chance of impacting Earth next century (Chesley et al., 2014; Farnocchia et al., 2021).

Here, we describe the delivery and distribution of the Bennu samples and present a first look at their physical and spectral characteristics, bulk elemental and isotopic composition, and mineralogy. We compare our findings to expectations from remote sensing and discuss the implications for Bennu's geologic history and relationship to other carbonaceous astromaterials.

### *Geologic Context of the Sample from Spacecraft Observations*

Corroborating the pre-launch telescopic observations (Clark et al., 2011; Lauretta et al., 2015), spectral data obtained by the OSIRIS-REx spacecraft showed that organic compounds and hydrated phyllosilicates are distributed ubiquitously on Bennu's surface (Hamilton et al., 2019; Simon et al., 2020a). Magnetite (Lauretta et al., 2019; Hamilton et al., 2019) and carbonates (Simon et al., 2020a,b; Kaplan et al., 2020) were also detected, and anhydrous silicates appear to be rare (Hamilton et al., 2021). These findings suggest that substantial aqueous alteration took place on a larger parent asteroid, estimated to have been 100–200 km in diameter (Lauretta et al., 2015, 2023a). This aligns with the hypothesis that Bennu is a rubble pile formed from the catastrophic disruption of its parent asteroid based on its dynamical association with inner-main-belt asteroid families (Walsh et al., 2013).

Bennu is among the darkest bodies in the solar system, with a median normal reflectance of 4.6% (Golish et al., 2021), consistent with its carbonaceous composition. Its surface is dominated by two major populations of boulders: "Dark" boulders have normal reflectance values of 3.5–4.9%, whereas "bright" boulders have normal reflectance values of 4.9–7.4% (DellaGiustina et al., 2020). Dark boulders often show signs of lamination and layering (Ishimaru and Lauretta, 2023). Some bright boulders have linear, meter-scale, higher-reflectance features, interpreted as carbonate veins resulting from late-stage hydrothermal alteration processes (Kaplan et al., 2020). Thermal inertia calculations indicate that bright boulders are stronger and denser overall than their darker counterparts, especially if they contain veins, though more nuanced variations in thermal inertia are observed at the local scale (Rozitis et al., 2020, 2022). Rare boulders with much higher reflectance (up to 26%) show spectral affinities with howardite-eucrite-diogenite (HED) and ordinary chondrite meteorites, indicating exogenic origins from asteroid (4) Vesta and stony asteroids, respectively (DellaGiustina et al., 2021; Le Corre et al., 2021; Tatsumi et al., 2021).

The OSIRIS-REx sample collection site, nicknamed Nightingale, is situated in Hokioi Crater (Figure 1B), a 20-m-diameter impact feature in Bennu's northern hemisphere (center coordinates approximately 56°N, 43°E). Like the global surface, Nightingale exhibits a spectral absorption band at 2.7 μm, indicative of hydrated phyllosilicates, and this is supported by a 23-μm feature in thermal emission spectra (Simon et al., 2020b; Hamilton et al., 2021). Additionally, the darkest material at this site shows a spectral absorption at 0.55 μm, suggesting the presence of magnetite ($Fe_3O_4$), which is further confirmed by features at 18 and 29 μm in thermal emission spectra (Simon et al., 2020b; Hamilton et al., 2021). Spectral evidence of carbon-bearing materials indicates carbonates with varying cation contents, including in veins in nearby boulders, and organic compounds with aromatic and aliphatic C–H bonds (Kaplan et al., 2020, 2021).

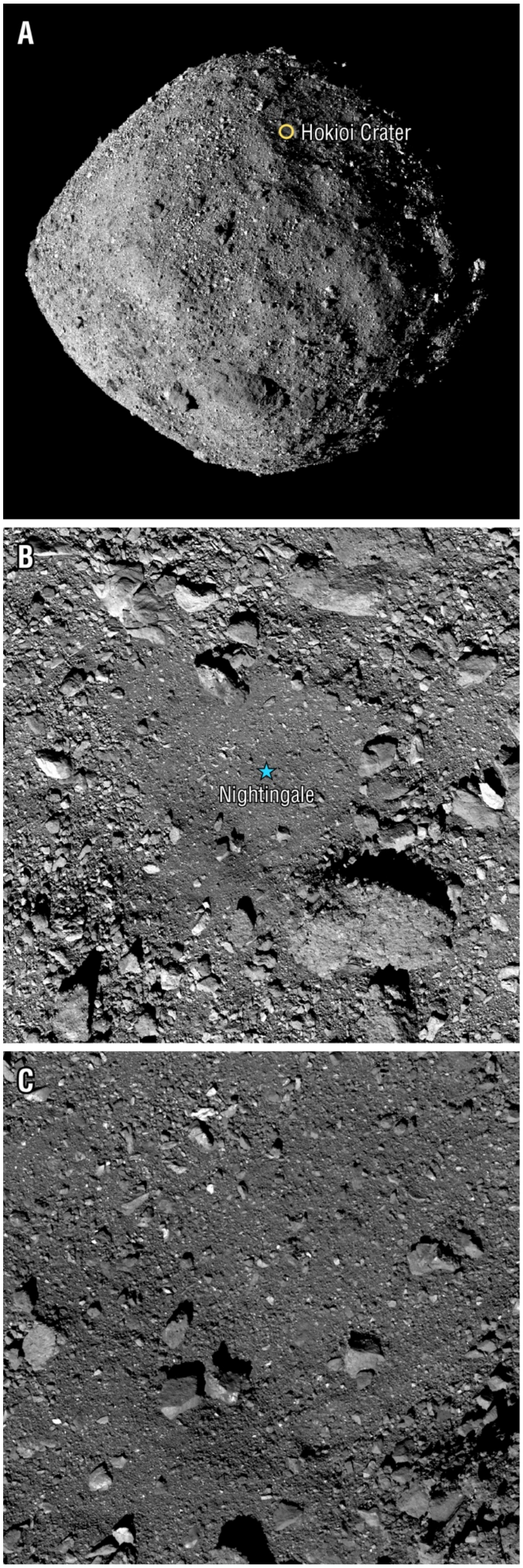
A
Hokioi Crater
B
Nightingale
C

**Figure 1: Bennu and the OSIRIS-REx sample collection site.** (A) Full-disk view of Bennu (equatorial diameter, ~500 m) from a range of 24 km. This mosaic consists of images collected using the PolyCam imager (Rizk et al., 2018; Golish et al., 2020) on 2 December 2018. Hokioi Crater (orange circle) is located near Bennu's north (+*z*) pole. (B) Mosaic of the 20-m-diameter Hokioi Crater, composed of PolyCam images collected on 26 October 2019 from a distance of about 1 km. The blue star indicates the Nightingale sample site. (C) Image of the Nightingale area taken by PolyCam on 26 October 2019. The field of view is 14.4 m. For reference, the lighter-colored boulder (far left middle) is 1.4 m long. [Image credits: NASA/Goddard/University of Arizona]

The regolith in Hokioi Crater has a salt-and-pepper appearance (Figure 1C) that suggests derivation from both dark and bright boulders (Lauretta et al., 2022; Walsh et al. 2022). Jawin et al. (2023) conducted a comprehensive investigation of boulder diversity in the broader region surrounding the Nightingale sample site, spanning longitudes 20°–80°E and latitudes 40°–70°N, using the high-resolution data acquired during the reconnaissance phase of the mission (Lauretta et al., 2021). This broad area is large enough to be representative of the diversity of the global surface. Dark boulders were subclassified into types A and B based on morphology: Type A boulders are characterized by rough, rounded, and clastic textures, whereas type B boulders exhibit intermediate roughness and angularity (Figure 2A–D). Resolvable clasts within type A boulders range from 10 to 120 cm in diameter. Some boulders display textures characteristic of both type A and B (Figure 2A), suggesting a potential genetic relationship between these two types. Bright boulders were subcategorized into types C and D (Figure 2E–G): Type C boulders are smooth, angular, and have flat planar faces, whereas type D boulders are slightly rougher and more rounded.

Hokioi Crater is posited to be one of Bennu's most recent impact features based on its small size and redder visible and near-infrared spectrum compared with the asteroid's blueish average surface (DellaGiustina et al., 2020; Barucci et al., 2020; Deshapriya et al., 2021). Its mid-latitude location experiences relatively moderate peak temperatures of around 360 K, compared with those in the equatorial region, which can exceed 390 K (Rozitis et al., 2022). These factors suggest that the regolith collected by OSIRIS-REx is among the freshest (least space-weathered) on Bennu's surface, though some particles may have migrated into the crater from higher latitudes via mass wasting (Barnouin et al., 2022; Jawin et al., 2020). Crucially for sample analysis, it is likely that the organic compounds at Nightingale have not been exposed for a duration significant enough to alter the H/C ratio or degrade aliphatic molecules (Kaplan et al., 2021).

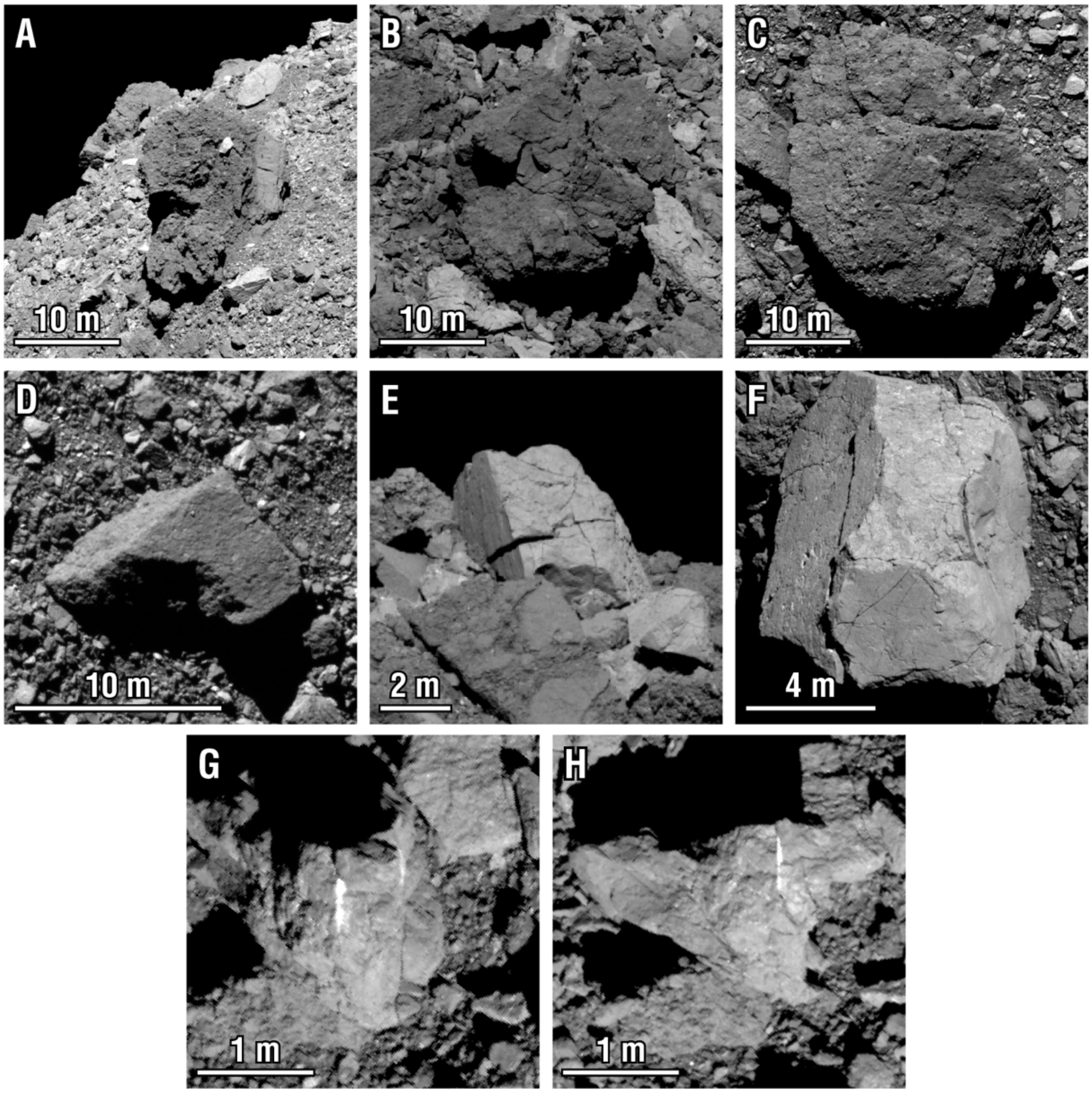

**Figure 2: Examples of boulder types on Bennu.** (A) In this boulder, called Gargoyle Saxum, most of the visible area exhibits the type A texture. The type B texture can also be seen at its base (upper right in this orientation). (B,C) Ciinkwia and Boobrie Saxa, both type A boulders. (D) Pouakai Saxum, a type B boulder. (E,F) Examples of type C boulders. (G,H) Two views of a type D boulder with high-reflectance veins. All images taken by PolyCam. [Image credits: NASA/Goddard/University of Arizona]

## Sample Collection and Delivery

The spacecraft obtained a sample of Bennu's regolith on 20 October 2020 using its Touch-and-Go Sample Acquisition Mechanism (TAGSAM), which comprises a specialized sampler head extended on a 3-m-long articulated arm (Bierhaus et al., 2018). TAGSAM collected two types of samples. The first was a bulk sample, wherein particles were directed into the TAGSAM head by the controlled release of high-purity nitrogen gas and downward penetration of TAGSAM ~0.5 m into the nearly cohesionless regolith (Lauretta et al., 2022; Walsh et al., 2022). During sampling, the TAGSAM head partially overlapped a ~20-cm-long, type A boulder with very low thermal inertia (possibly indicating low density) that may have been crushed and therefore sampled (Lauretta et al., 2022; Walsh et al., 2022; Ryan et al., 2024). The second type of sample consisted of particles captured in 24 “contact pads” made of stainless-steel Velcro® on the TAGSAM baseplate. The contact pads were designed to trap fine particles and dust directly from the optically active surface layer. Spacecraft images of TAGSAM after sampling indicated that the bulk sample encompasses a range of particle sizes, from sub-micron dust to fragments measuring up to ~3 cm in diameter, and that the contact pads retained millimeter-sized particles (Lauretta et al., 2022).

On 10 May 2021, the OSIRIS-REx spacecraft departed Bennu and embarked on a trajectory that would intersect Earth’s orbit more than two years later. The spacecraft released the Sample Return Capsule (SRC) early in the morning of 24 September 2023 with a spin of 13.27 rpm and a relative velocity of 31.3–33.6 cm/s, both parameters within the design range (Getzandanner et al., 2024). After the SRC release, the spacecraft executed a divert maneuver that put it on a trajectory for its extended mission as OSIRIS-APEX (OSIRIS–Apophis Explorer), en route to a rendezvous with asteroid (99942) Apophis in April 2029 (DellaGiustina et al., 2023).

The capsule reached the atmospheric interface at 14:42 UTC near the California coast. From that point forward, the flight of the SRC was captured by visible and infrared video footage, which suggested a nominal entry through the hypersonic phase (Francis et al., 2024). The peak heating was estimated to occur within 2 seconds of the predicted time, as indicated by the intensity in the infrared video. However, the drogue parachute deployed more than 3 minutes late, immediately followed by the main parachute. The capsule was oscillating or tumbling before the drogue deployed. Despite this deviation from the expected sequence, the main parachute deployment and descent were nominal (Figure 3A), leading to a safe touchdown about 3 minutes earlier than predicted at 14:52 UTC at the Utah Test and Training Range (UTTR) in Dugway, Utah (Figure 3B,C).

The recovery team reached the capsule landing site via helicopter within 20 minutes of touchdown (Figure 3D,E). After ensuring safety measures and covering the vents and parachute cavity, they wrapped the 50-kg capsule in layers of polytetrafluoroethylene and a tarp (Figure 3F). Using a long line, the capsule was airlifted by helicopter to a nearby temporary cleanroom (Figure 3G). There, a team in protective gear carefully disassembled the capsule, extracting the science canister containing the TAGSAM head and sample (Figure 3H). A high-purity $N_2$ purge was applied to the science canister to displace any air that might have entered through the canister filter, creating an environment conducive to long-term safeguarding against terrestrial contamination. Any loose particles were collected for future assessment as potential asteroidal material.

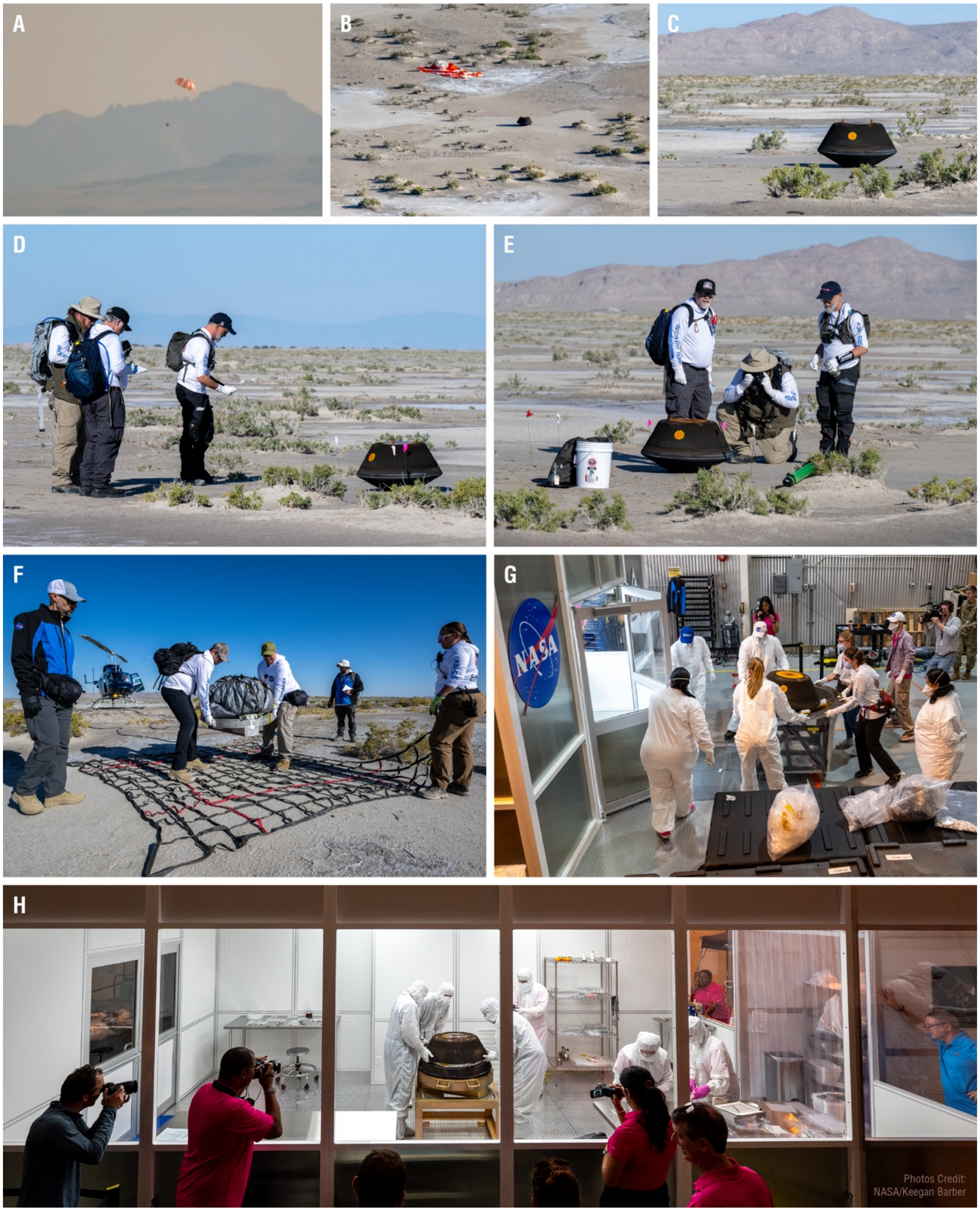

**Figure 3: Descent, recovery, and transport of the capsule.** (A) The capsule and main parachute during the final stages of descent. (B) The capsule on the ground shortly after landing, with the main chute visible in the background. (C) A close-up view of the capsule, with an orange patch placed by safety personnel to cover a vent in the backshell. (D,E) Members of the recovery team documenting the capsule's state and the local environment. (F) The capsule being bagged and placed in a cargo net for transport to a nearby facility. (G) The capsule being unbagged and quickly moved into a portable cleanroom at UTTR. (H) Inside the temporary cleanroom, the capsule being partially disassembled and prepared for transport to JSC. [Image credits: NASA]

The next morning (25 September 2023), the science canister was transported under continuous $N_2$ purge via a U.S. Air Force C-17 aircraft to Ellington Field, Houston, Texas, then via delivery truck to NASA's Johnson Space Center (JSC). That afternoon, the science canister was transferred into a custom glovebox in the OSIRIS-REx cleanroom (Righter et al., 2023).

When the canister was opened on 26 September 2023, dark powder and sand-sized particles that had escaped the TAGSAM head were observed on the avionics deck and adhering to the exterior of TAGSAM and the canister lid (Figure 4) — an expected scenario that enabled rapid initial characterization. This material was concentrated at the canister lid interface, probably because of the tumbling of the SRC or the outward pressure of the purge gas.

The curation team then removed 14 circular witness plates from the top of the TAGSAM head. These plates had been monitoring the environmental conditions of the canister and the TAGSAM head at different stages of the mission (Dworkin et al., 2018). Some of the plates showed visible dust (Figure 4), likely from Bennu, whereas others had been covered at specific operational milestones to help researchers identify potential contamination sources. All witness plates and associated hardware were securely stored as part of the OSIRIS-REx contamination knowledge collection.

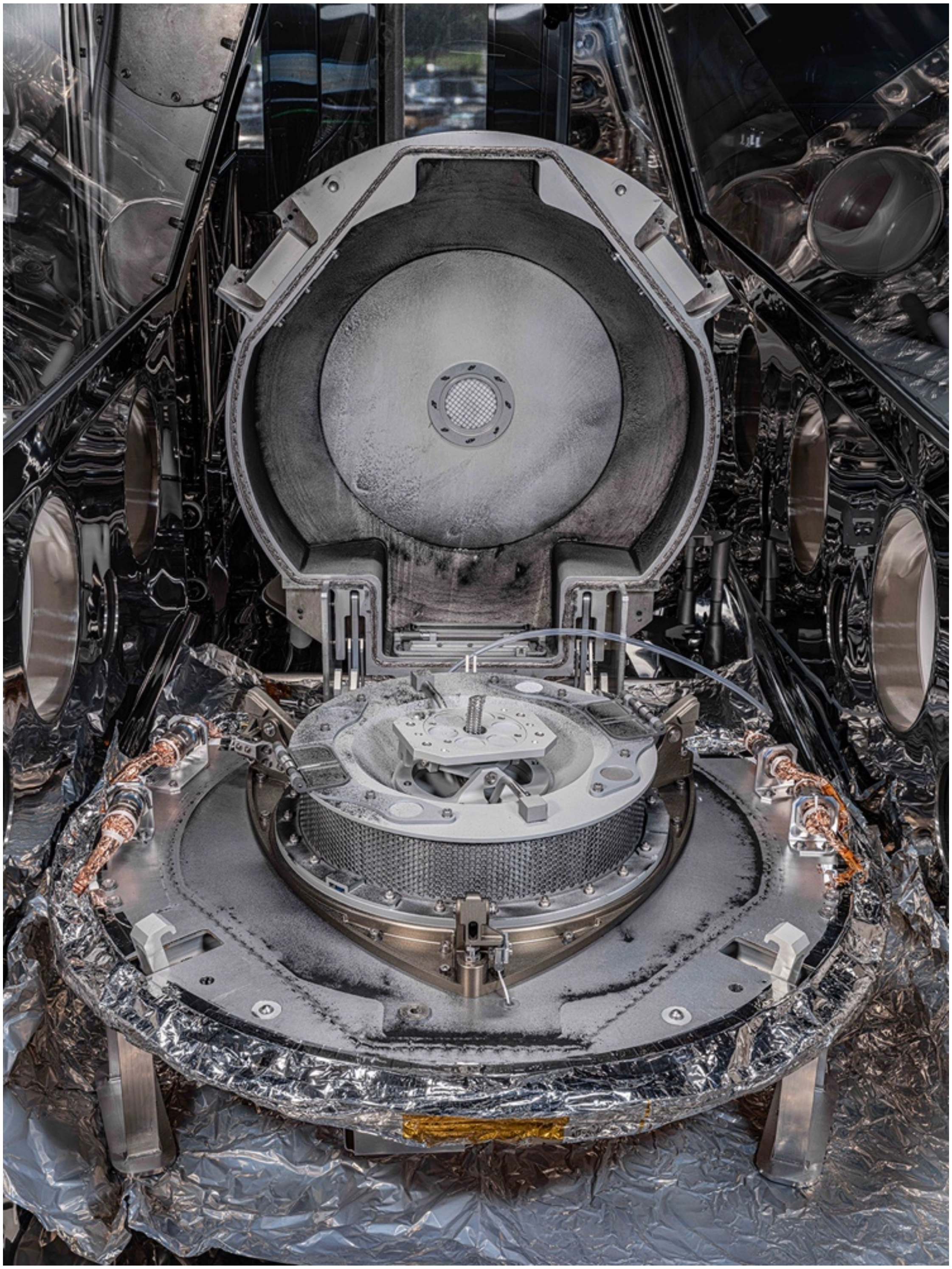

**Figure 4: The opened science canister in the glovebox at JSC.** The cylindrical TAGSAM head is positioned in the center of the avionics deck. Dark particles and dust can be seen on the avionics deck, the top of TAGSAM, and the inside of the canister's lid.

Next, the TAGSAM head was detached from the avionics deck and flipped over to reveal the baseplate, including the 24 contact pads. This position enabled access to the sample inside.

In the weeks that followed, curation scientists and technicians embarked on the delicate task of disassembling the TAGSAM head, which was transferred to its own glovebox for processing. Both gloveboxes are continuously purged with high-purity $N_2$ to preserve the sample's pristine condition.

Two of the 35 fasteners securing the TAGSAM head could not be unscrewed using the approved tools available in the glovebox, which initially hindered its complete disassembly. In response, the team developed a novel strategy to extract regolith from the head while preserving its pristine condition. First, curators collected particles from the top of TAGSAM's Mylar flap, distributing them among two specially designed wedge-shaped trays (Figure 5A, left). Subsequently, by holding down the Mylar flap and utilizing forceps or a scoop, they retrieved regolith from the interior of TAGSAM, distributing it among two additional wedge-shaped trays (Figure 5A, right). At this point, 70.3 g of rocks and dust had been obtained from the Mylar flap, the TAGSAM head, and the avionics deck, exceeding the mission requirement of 60 g (Lauretta et al., 2017).

However, additional material was still present within TAGSAM. In mid-January 2024, the curation team successfully removed the two stuck fasteners using a custom tool. They poured the remaining sample into eight wedge-shaped trays (Figure 5B). An additional 51.8 g was obtained during this step, leading to a grand total of at least 121.6 g of Bennu sample delivered to Earth by OSIRIS-REx — less than the 250 ± 101 g predicted from spacecraft data (Ma et al., 2021; Lauretta et al., 2022), but well above the mission requirement.

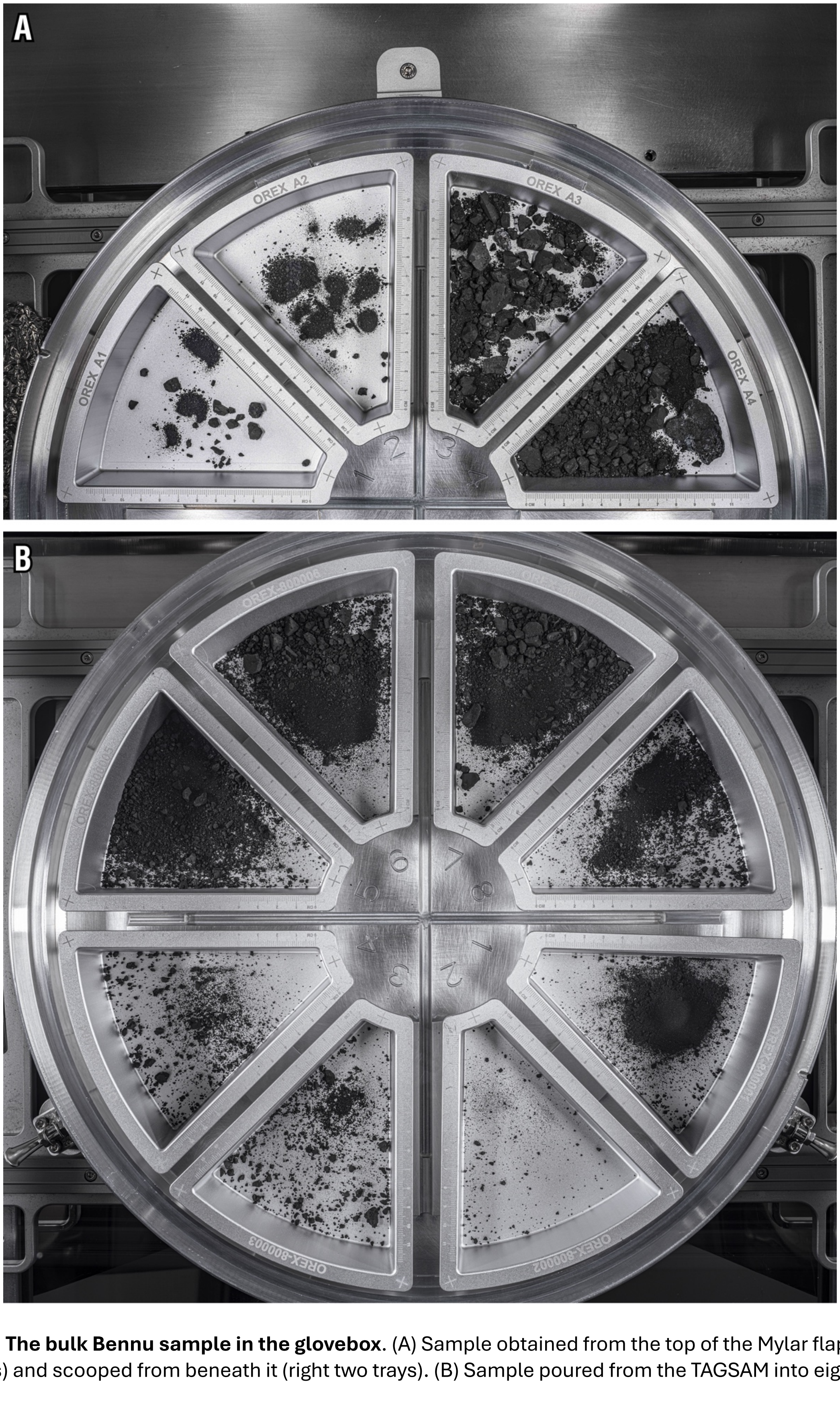

**Figure 5: The bulk Bennu sample in the glovebox**. (A) Sample obtained from the top of the Mylar flap (left two trays) and scooped from beneath it (right two trays). (B) Sample poured from the TAGSAM into eight trays.

# Methods and Results

## Sample Distribution

The Bennu regolith present on the avionics deck was sampled on 27–28 September 2023 for "quick-look" (QL) analyses by the OSIRIS-REx Sample Analysis Team (SAT) (Lauretta et al., 2023b). The QL allocation included aggregate samples composed of fine (<100 μm), intermediate (100–500 μm), and coarse (500 μm–5 mm) particles. Smaller particles displayed a tendency to clump. This material was distributed among four samples (Table 1).

The QL samples contained numerous straight and curving fibers, likely spacecraft contaminants from the multi-layer insulation and batting in the SRC, interspersed throughout the fines. The insulation was cut and vacuumed as part of the recovery operations at the UTTR, but any material that was trapped under the canister lid's outer lip could have a path to the avionics deck. The fibers were manually removed with forceps by inspection under a reflected-light microscope before further subsampling and analysis.

In November 2023, an initial set of eight aggregate samples, ranging from fine to coarse particles, of the material manually retrieved from TAGSAM (Figure 5A) was distributed among the SAT (Table 1). Lastly, 12 large stones (>5 mm; Table 1) of various macroscopic appearances were removed from the trays for detailed characterization within the glovebox.

**Table 1: Samples analyzed in this study.**

| Sample ID | Mass (mg) | Collection location | Type | Techniques* (sub-sample numbers) |
|---|---|---|---|---|
| OREX-500002-0 | 22 | Avionics deck | Aggregate | EA-IRMS (OREX-501033-0, OREX-501034-0, OREX-501035-0, OREX-501036-0, OREX-501037-0, OREX-501038-0, OREX-501039-0, OREX-501040-0, OREX-501041-0) |
| OREX-500003-0 | 12 | Avionics deck | Aggregate | XCT (OREX-500003-100); SEM/EDS (OREX-501002-0, OREX-501017-0); UV-L2MS-μFTIR (OREX-501006-0) |
| OREX-500005-0 | 88 | Avionics deck | Aggregate | XRD (OREX-500005-0); SEM/EDS (OREX-501001-0); UV-L2MS (OREX-501018-0); μFTIR (OREX-501000-0); ICP-MS (OREX-501043-0); LF (OREX-501042-0, OREX-501047-0, OREX-501066-0, OREX-501067-0); NanoSIMS (501018-100) |
| OREX-500007-0 | 16 | Avionics deck | Aggregate | SEM/EDS (OREX-501004-0) |
| OREX-800014-0 | 6236 | TAGSAM | Stone | SLS |
| OREX-800017-0 | 2042 | TAGSAM | Stone | SLS |
| OREX-800019-0 | 2251 | TAGSAM | Stone | SLS |
| OREX-800020-0 | 576 | TAGSAM | Stone | SLS |
| OREX-800021-0 | 592 | TAGSAM | Stone | SLS |
| OREX-800023-0 | 399 | TAGSAM | Stone | SLS; allocated to SAT |
| OREX-800026-0 | 250 | TAGSAM | Stone | SLS |
| OREX-800028-0 | 148 | TAGSAM | Aggregate | SEM/EDS (OREX-803009-101) |
| OREX-800029-0 | 201 | TAGSAM | Aggregate | RELAB (OREX-800029-0) |
| OREX-800031-0 | 52 | TAGSAM | Aggregate | EA-IRMS (OREX-803040-0, OREX-803041-0, OREX-803042-0, OREX-803043-0, OREX-803002-0, OREX-803044-0, OREX-803045-0, OREX-803046-0) |
| OREX-800033-0 | 131 | TAGSAM | Aggregate | ICP-MS (OREX-803015-0); SEM/EDS (OREX-803079-0, OREX-803080-0, OREX-803100-0) |
| OREX-800055-0 | 591 | TAGSAM | Stone | SLS; allocated to SAT |
| OREX-800067-0 | 296 | TAGSAM | Stone | SLS |
| OREX-800087-0 | 372 | TAGSAM | Stone | SLS |
| OREX-800088-0 | 329 | TAGSAM | Stone | SLS; allocated to SAT |
| OREX-800089-0 | 313 | TAGSAM | Stone | SLS |

*Abbreviated technique names are defined in the text.

## Particle Morphologies and Physical Properties

The Advanced Imaging and Visualization of Astromaterials (AIVA) procedure was used for comprehensive imaging to document the condition of the hardware upon arrival, the disassembly process, and the visual characteristics of the sample.

AIVA imagery is created using a combination of manual high-resolution precision photography and a semi-automated focus stacking process. For the former, we use a medium-format camera system with a full-frame 102-megapixel sensor and a variety of lenses. The sample remains inside its glovebox, which is constructed from highly reflective materials, and all camera and lighting equipment are positioned outside the glovebox. This setup limits our range of motion around the sample, requiring us to essentially perform macro photography at telephoto distances through small glass windows into a mirrored box.

Our focus stacking technique involves capturing hundreds of images, each with a tiny area in focus within the field of view, to achieve sharp focus detail throughout an entire image. During post-imaging processing, we work with the full stacks simultaneously in RAW format, manually adjusting brightness, density, color, and other settings to achieve a balanced image. We use an automated blending algorithm to process the image stack, making additional manual adjustments as needed to reduce processing artifacts and reveal obscured data.

Using the AIVA imaging process, we captured initial views of the avionics deck with the TAGSAM and witness plates just after the canister was opened (Figure 4). Additionally, detailed imaging was performed of the interior of the canister lid, the avionics deck, the base of TAGSAM, and the capture ring that had held TAGSAM in place. Special attention was given to the sample after it was distributed in the 12 wedge-shaped trays (Figure 5). A series of images was taken at this point, including zoomed views of each of the 12 trays, providing essential data for the initial characterization of the appearance and size distribution of the bulk sample.

### *Apparent Brightness and Morphologic Properties*

The AIVA images showed that most of the sampled material is very dark, although many particles have high-reflectance inclusions distributed in the dark groundmass. This texture is apparent in the fine, intermediate, coarse, and large particles. Some of the high-reflectance phases have a hexagonal crystal habit, whereas others appear as clusters of small spheres, platelets, and dodecahedral forms. In addition, highly reflective individual particles occur throughout the collection. Visual analysis of AIVA images indicates that particles in the sample can be broadly grouped into three categories based on morphology and apparent brightness:

1. **Hummocky** (Figure 6A–F): These particles exhibit a sub-rounded shape with an uneven surface, featuring small, rounded mounds and depressions reminiscent of cauliflower. This texture gives the stones a lumpy or "hummocky" appearance, with varying sizes and shapes of the mounds. These particles are generally dark, but occasionally include small (tens to hundreds of microns), higher-reflectance phases dispersed throughout.
2. **Angular** (Figure 6G–K): These particles stand out owing to their fracturing, which is characterized by sharp angular faces, resulting in hexagonal or polygonal shapes. These particles have straight and parallel faces, giving them a distinct geometric appearance. Some show apparent layering in the form of step-like parallel exposed faces. Though generally dark, some angular particles exhibit specular reflections or metallic luster on individual faces. Highly reflective inclusions, similar in appearance to those in the hummocky stones, are present throughout these samples.
3. **Mottled** (Figure 6L–P): These particles exhibit higher-reflectance material that either adheres to, encases, or overlays the low-reflectance material, which makes up the bulk of the stones. Additionally, the higher-reflectance material appears to fill cracks as veins that are tens of microns thick. This higher-reflectance material also occurs as isolated bright flakes throughout the fine component, suggesting its friable nature and ease of separation from the host rock.

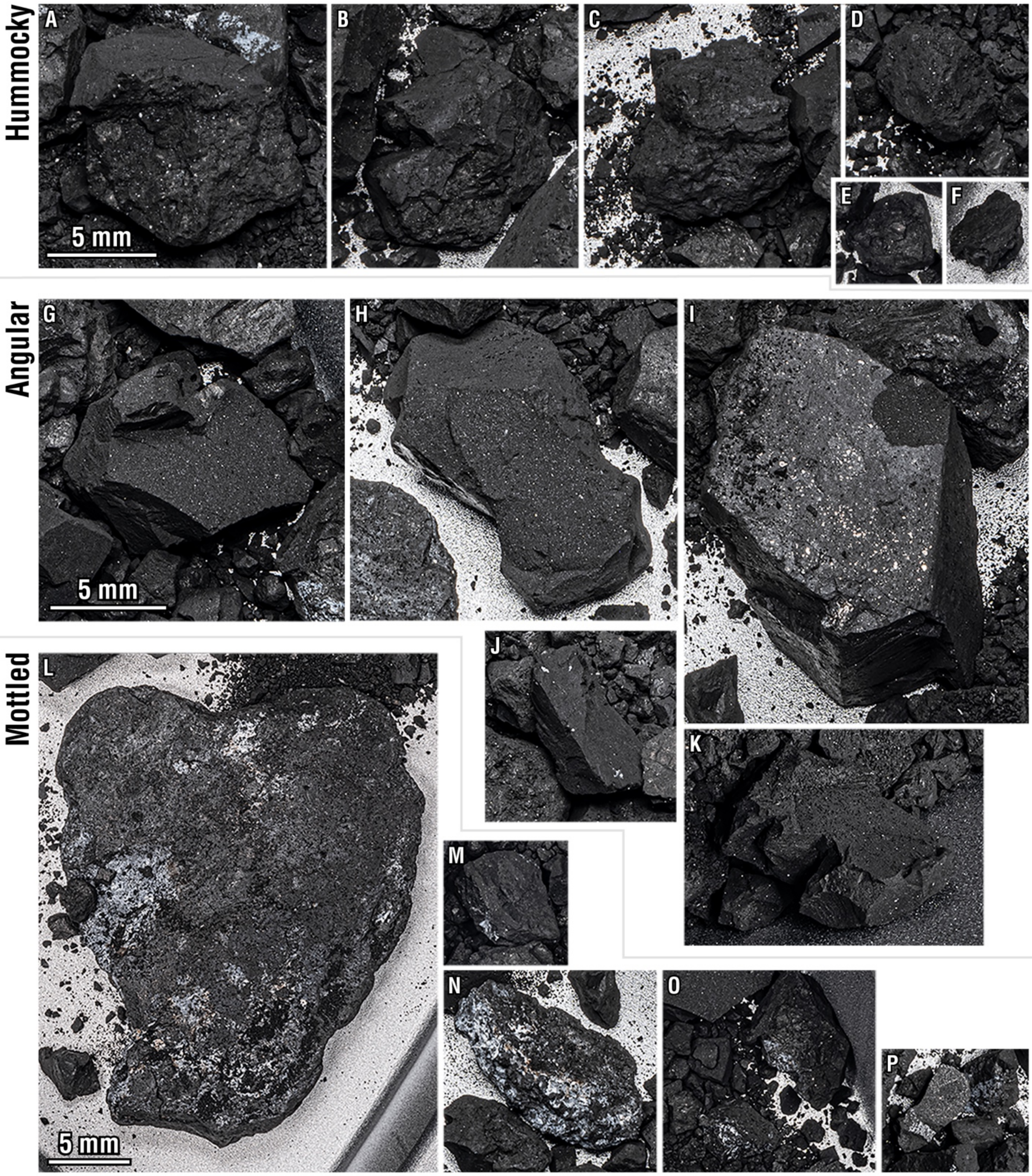

**Figure 6: Three broad categories of particles observed in the sample.** AIVA images showing examples of hummocky (A–F), angular (G–K), and mottled (L–P) particles.

*Size Frequency Distribution*

The particle size frequency distribution (PSFD) of the sample in the trays was measured at the Astronomical Observatory of Padova using the AIVA images. The longest axis of each identifiable particle in the photos was manually mapped as a line segment in the ArcGIS software package. The largest particles, which are naturally less abundant, were counted first. Smaller particles were subsequently counted until the remaining particles were too small to be adequately resolved across five pixels.

A total of 7154 particles were identified across all 12 trays (Figure 7): 2220 particles in the first four trays (sample scooped or picked from TAGSAM) and 4934 particles in the eight additional trays (sample poured out of TAGSAM). About 95% (6788) of the counted particles are at least 0.5 mm in length, nearly two-thirds (4438) are at least 1.0 mm, and about one-third (224) are at least 5.0 mm. Most of the counted particles (6565, or 92%) fall between 0.5 and 5.0 mm. Larger particles are less common, with only 34 particles at least 10 mm, seven particles at least 15 mm, and two particles at least 20 mm. The largest particle measured — an instance of the mottled type — is about 35 mm. The median size is 1.2 mm and the mean is 1.6 mm. The standard deviation about the mean is 1.5 mm.

The energetic process of collection and recovery likely fractured some of the material (Lauretta et al., 2023b). Therefore, the observed PSFD represents only a limit on the original distribution prior to collection.

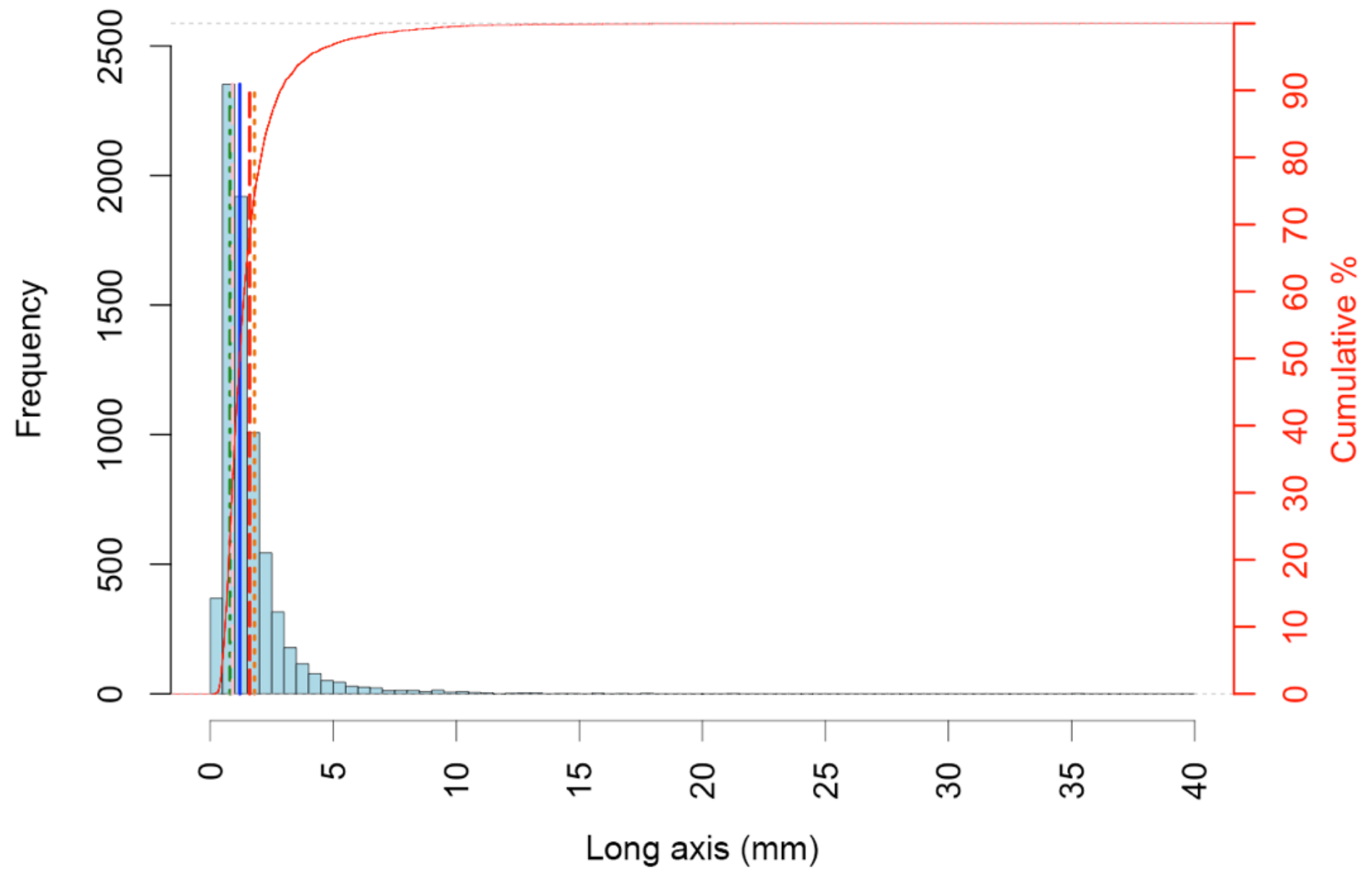

**Figure 7: Size frequency distribution of >7000 particles measured in images of the bulk sample in trays.** The first and third quartiles are indicated by the green-dashed and orange-dotted vertical lines. The dashed pink line is the mode, and the solid blue line is the median and the red dashed line is the mean. The distribution extends to the largest particle at 35 mm.

### *Interior Structure*

We used the Nikon XTH 320 X-ray computed tomography (XCT) instrument at JSC to scan an angular particle measuring 2.0 × 1.8 × 0.8 mm and weighing 1.96 ± 0.01 mg (Table 1). Its appearance is typical of most of the collection, exhibiting a dark, fine-grained groundmass with embedded euhedral and subhedral metallic-luster phases. This particle was positioned within a polyethylene pipette tip, and radiographs were acquired, enabling detailed visualization of its internal structure. X-ray conditions were set at 80 kV and 0.036 mA, and a total of 3141 radiographs were collected at an exposure time of 4.00 seconds per radiograph as the sample rotated 360°. This comprehensive scan took about 3.5 hours to complete. Radiographs were subsequently reconstructed into a three-dimensional (3D) volume with a voxel size of 2 μm per voxel edge. Nikon's CTPro3D (v. 5.4) software was employed for this reconstruction, and the resulting data were exported as a series of contiguous 16-bit grayscale TIFF images. In these images, brighter grayscale values generally correspond to denser materials.

The particle's volume was determined through four measurements using different data segmentation techniques. These measurements yielded a range of 1.15 to 1.16 $mm^3$. Macroporosity remains challenging to unambiguously identify, primarily owing to numerous spherical and dark regions that are almost exclusively in contact with high-atomic-number materials. Many of these dark regions are probably shadows created by X-ray scattering.

The groundmass primarily comprises a single, fine-grained, low–X-ray–attenuating material (Figure 8A). Within this material, slightly brighter (more X-ray–attenuating) regions appear as three-dimensionally coherent domains with boundaries ranging from sharp to diffuse. This brighter material exhibits relative homogeneity and, in certain instances, forms planar to bladed structures measuring approximately 450 μm in length by 30 μm in thickness (Figure 8).

The high-atomic-number (e.g., Fe-bearing) materials manifest in various sizes, shapes, and grayscale brightness levels. Collectively, they constitute 7.1 ± 1.8% (1σ) of the particle volume. Uncertainties stem from utilizing distinct segmentation methods for measuring the entire particle volume and the volume occupied by high-atomic-number materials.

A subset of the high-density crystals has distinctive hexagonal habit, size, and shape (Figure 8B–E). The average long, intermediate, and short axes of these crystals are 23, 17, and 12 μm, respectively. Long axes exhibit an average length 1.3 times greater than the intermediate axis and 1.9 times greater than the short axis. The largest crystal identified is 134 × 84 × 59 μm.

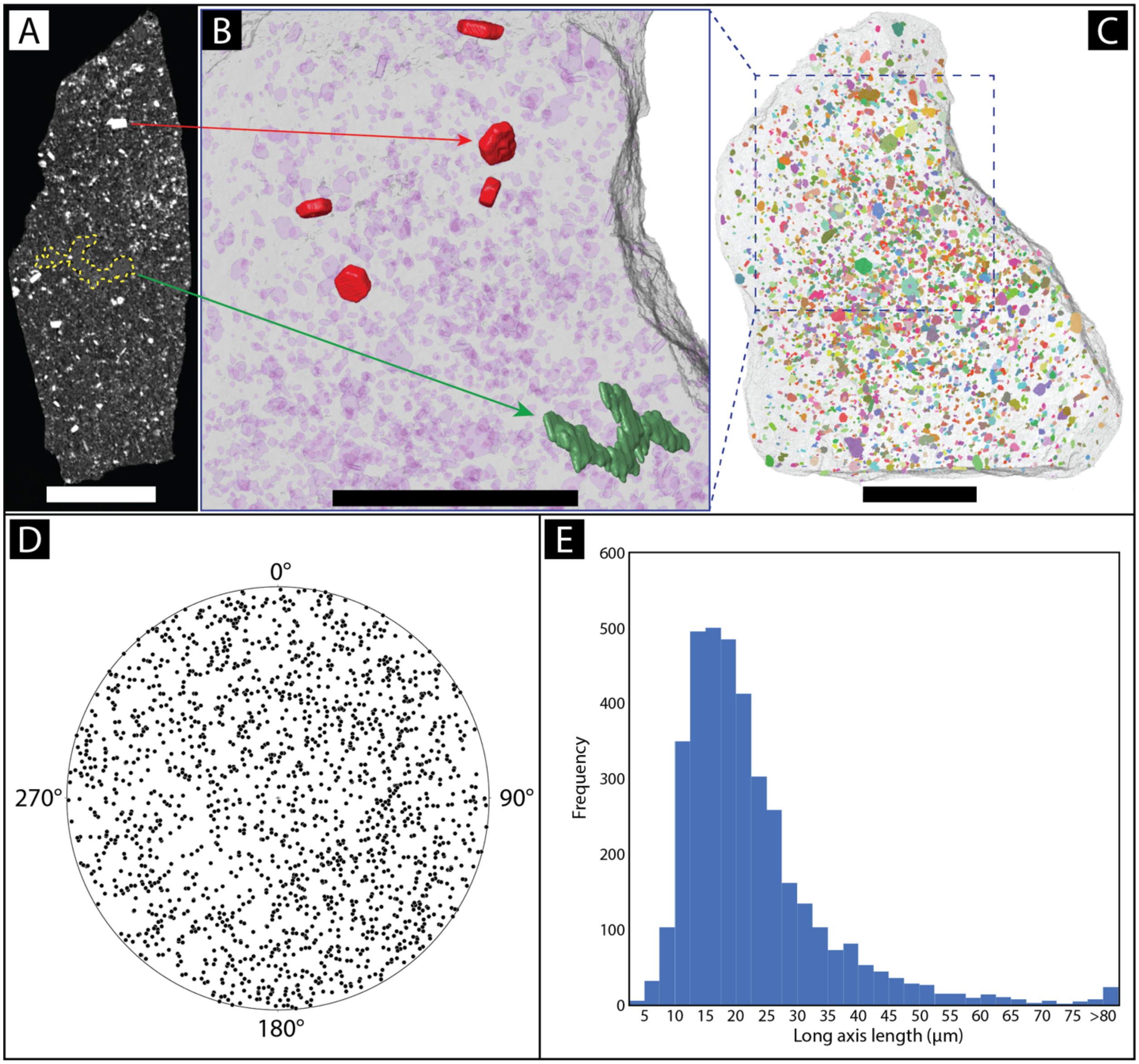

**Figure 8: XCT analysis of a representative particle.** (A) A 16-bit grayscale XCT slice of OREX-500003-100 displaying hexagonal crystals as the brightest objects. The red arrow highlights the largest single crystal. The dark groundmass appears mostly homogeneous with occasional slightly brighter regions. The largest instance of this slightly brighter material is outlined in yellow. (B) Zoomed-in 3D rendering illustrating the exterior surface (transparent gray) with all segmented hexagonal crystals (transparent purple) and some of the largest grains (red) to emphasize their hexagonal nature. The green object corresponds to the slightly brighter material outlined in yellow in (A). (C) 3D volume rendering of the entire sample (transparent gray) with individual labeling of all segmented hexagonal crystals. (D) Stereonet displaying the short axis orientation of crystals with a short axis length exceeding 4 voxels (8 μm). (E) Histogram illustrating the distribution of crystal long axis lengths with bin sizes of 2.5 μm. Scale bars in A, B, and C are 0.5 mm.

### *Density*

The shapes and volumes of five hummocky, five angular, and two mottled stones (Table 1, 2) were measured individually using a Polyga Compact C506 structured light scanner (SLS). This system uses a blue-light LED projector to cast parallel stripes of light onto the stone through the windows of the glovebox. Positioned on either side of the projector are two cameras that capture the stone from different angles to measure the distortion of the light patterns as they contour to its 3D form. The stone is rotated 10–20 times to obtain a partial 3D model of its top half. After flipping the stone, the scanning and rotating process is repeated for the other half. The partial scan files collected from each perspective are merged using the FlexScan3D software to create a full 3D model of the object, at which time any remaining holes are filled to create a watertight model.

We used the measured masses of the stones and their external volumes calculated from the 3D models to determine their bulk density (Table 2). Volume accuracy is ≤1% based on test measurements of irregularly shaped aluminum particles of known volume.

The hummocky stones are the least dense, with a weighted average of 1.55 ± 0.07 g/cm$^3$ from five stones with a combined mass of 1.856 g. The angular stones have higher densities, averaging 1.69 ± 0.04 g/cm$^3$ from five stones (combined mass = 5.756 g). The mottled stones show the highest measured densities, averaging 1.77 ± 0.04 g/cm$^3$ from two stones with a combined mass of 6.635 g (including the largest stone in the collection, with a mass of 6.236 g). Combined, these measured stones represent 11.7% of the total 121.6 g collection.

**Table 2: Masses, volumes, and densities of 12 stones.**

| Sample ID | Type | Mass (g) | SLS volume (cm$^3$) | Density (g/cm$^3$) | Density uncertainty |
|---|---|---|---|---|---|
| OREX-800089-0 | Hummocky | 0.313 | 0.2109 | 1.48 | 0.03 |
| OREX-800087-0 | Hummocky | 0.372 | 0.2476 | 1.50 | 0.03 |
| OREX-800026-0 | Hummocky | 0.250 | 0.1624 | 1.54 | 0.03 |
| OREX-800021-0 | Hummocky | 0.592 | 0.3780 | 1.57 | 0.02 |
| OREX-800088-0 | Hummocky | 0.329 | 0.2004 | 1.64 | 0.03 |
| **Hummocky weighted average density** | | | | **1.55** | **0.07** |
| OREX-800067-0 | Angular | 0.296 | 0.1847 | 1.60 | 0.03 |
| OREX-800020-0 | Angular | 0.576 | 0.3462 | 1.66 | 0.02 |
| OREX-800019-0 | Angular | 2.251 | 1.3329 | 1.69 | 0.02 |
| OREX-800017-0 | Angular | 2.042 | 1.2084 | 1.69 | 0.02 |
| OREX-800055-0 | Angular | 0.591 | 0.3390 | 1.74 | 0.02 |
| **Angular weighted average density** | | | | **1.69** | **0.04** |
| OREX-800014-0 | Mottled | 6.236 | 3.5408 | 1.76 | 0.02 |
| OREX-800023-0 | Mottled | 0.399 | 0.2183 | 1.83 | 0.03 |
| | | | **Mottled weighted average density** | **1.77** | **0.04** |

## Spectral Properties

An aggregate sample from the interior of TAGSAM (Table 1) was spectrally characterized at the NASA Reflectance Experiment Laboratory (RELAB) at Brown University. Upon arrival, the sample was extracted from the shipping container and placed on weighing paper and a transfer dish within a standard black Teflon-coated measurement dish (14 mm diameter). Residual fine particles adhering to the weighing paper and transfer dish were removed and mounted separately in a smaller Teflon-coated measurement dish (2.5 mm diameter).

Reflectance spectra were acquired with a custom bi-directional spectrometer (BDR) and Nexus/Nicolet 870 Fourier transform infrared (FTIR) spectrometer (Figure 9). The BDR measured wavelengths in the range of 0.30–2.6 μm for both splits, whereas the FTIR measured wavelengths of 0.8–100 μm (12,500–100 $cm^{-1}$) for the larger split and 0.8–25 μm (12,500–400 $cm^{-1}$) for the smaller sample. Visual inspection of the smaller split revealed bright particles. To assess their composition, isolated spectra of these bright particles were acquired using a Bruker microscope FTIR. Subsequently, the bright particles were manually removed and mounted on a scanning electron microscopy (SEM) stub for separate analyses. The rest of the two splits were then reintegrated for further investigations.

The spectra of the two splits have a similar, generally “red” spectral slope from 0.50 to 1.3 μm (Figure 9). There are variations in the smaller sample’s spectrum observed beyond 1.3 μm, potentially attributable to sample size disparities or the presence of bright particles. Both splits feature a broad but shallow absorption at 0.90–1.5 μm. The spectrum of the larger sample is dominated by a sharp absorption feature at 2.7 μm, consistent with Mg–OH vibrations. Additional absorption features are centered at 3.4 μm and 3.8–4 μm, consistent with the presence of carbonates and organic molecules. The bulk spectral properties of the measured sample closely resemble laboratory spectra of Ryugu samples acquired at RELAB (Hiroi et al., 2023). Reflectance properties in the near to mid-IR region are also consistent with those of some previously studied CI chondrites (Amano et al., 2023).

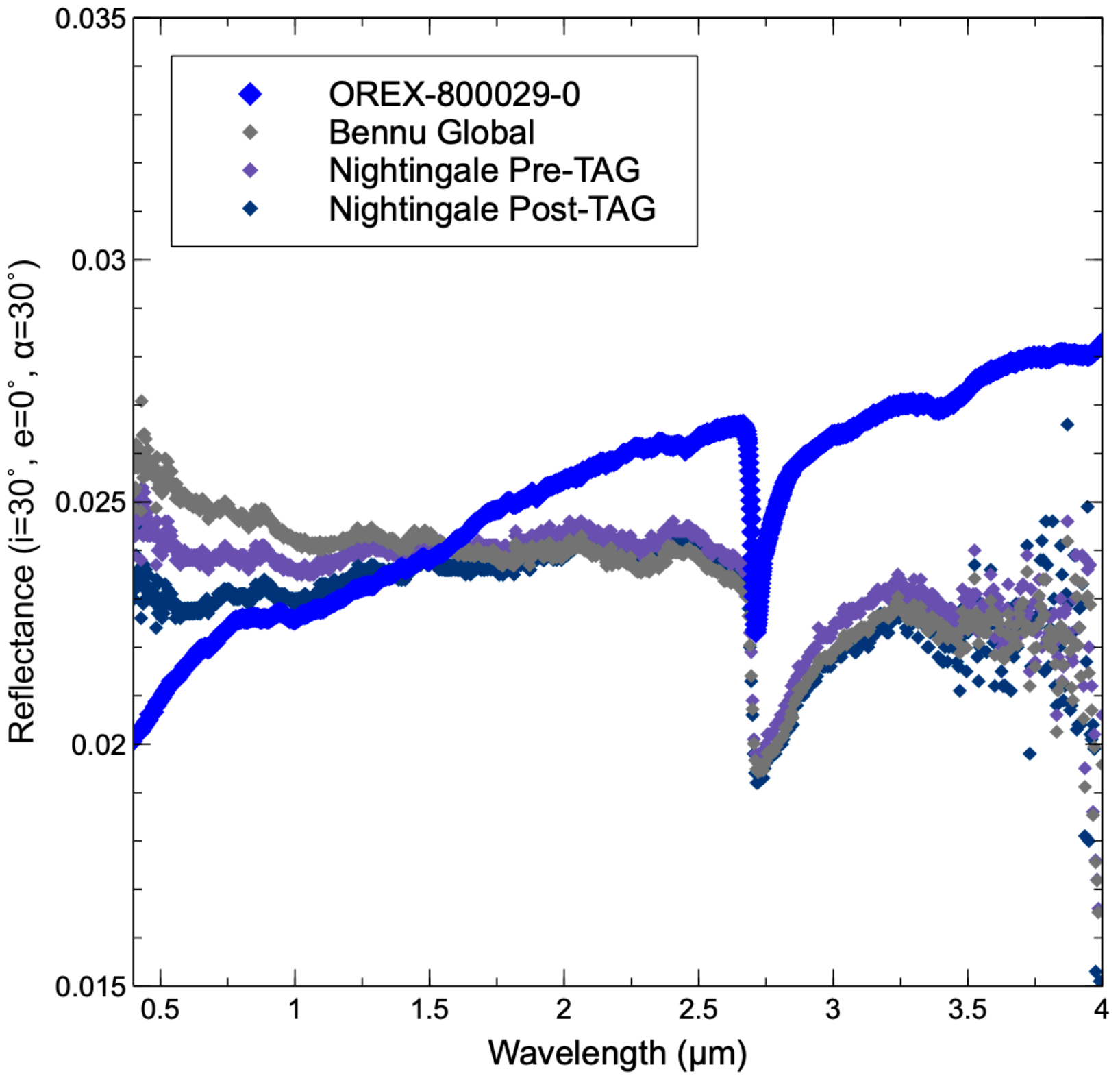

**Figure 9: OVIRS spectra of the asteroid compared with the RELAB spectrum of a sample.** Spectra, in order of bluest to reddest slope, of global Bennu, Nightingale before sampling (TAG), Nightingale after sampling, and a returned sample. OVIRS spectra of Bennu and the Nightingale sample site were corrected to the RELAB standard observing conditions (incidence = 30˚, emission = 0˚, and phase angle = 30˚).

The spectra were corrected for photometric effects to allow for a direct comparison with data collected at Bennu by the OSIRIS-REx Visible and InfraRed Spectrometer (OVIRS) (Reuter et al., 2018, Simon et al., 2021). The spectral shape of the 2.7-μm feature measured at RELAB is weaker and narrower than in the OVIRS data from Bennu, both for the global average and the Nightingale sample site. The absence of a strong feature in the broader 3-μm region indicates that the sample may be depleted in absorbed $H_2O$, relative to the surface of Bennu.

The sample's brightness matches that of the global asteroid around 1.1–1.2 μm. However, their spectral slopes differ considerably from 0.4 to 2.5 μm, with the analyzed sample having a more positive (redder) slope (Figure 9). Data from OVIRS and the MapCam multicolor imager (Rizk et al., 2018; Golish et al., 2020) showed an increase in spectral slope at Nightingale after sampling (Lauretta et al., 2022). The sample analyzed at RELAB is even redder than the post-sampling surface. If this sample represents the average composition of material at the Nightingale site, there must be other differences between the returned sample and the asteroid surface — such as particle size, rock surface texture, or the degree of space weathering — to account for the difference in spectral slopes. This difference may indicate that some of the sample was collected from depths down to 48.8 cm, owing to the compliant nature of the surface (Lauretta et al., 2022).

## Bulk Elemental and Isotopic Abundances

### *Bulk Elemental Abundances*

Major and trace elemental analyses were carried out using a Thermo Fisher iCAP Qc inductively coupled plasma mass spectrometer (ICP-MS) at Washington University in St. Louis. We analyzed the bulk major and trace elemental compositions of a QL sample from the avionics deck and a split of a sample from the interior of TAGSAM (Table 1). The samples were dissolved using a mixture of concentrated HF and $HNO_3$ at a 3:1 ratio, with additional treatments using HCl and $H_2O_2$ to remove fluorides and organics, respectively. Linear calibrations were performed using a synthetic CM chondrite standard at multiple dilution factors, with the standard reference materials BIR-1 and BHVO-2 run alongside to monitor the standard's performance. A 5-ppb internal standard of Re + Rh was consistently employed throughout the analysis to correct for instrument drift. The samples were run three times each, with total dilution factors of about 5000 and 50,000 for major and trace element data, respectively.

To ensure data quality and facilitate direct comparisons, nine carbonaceous chondrite fall samples, ranging from 14 to 69 mg, were also analyzed alongside the Bennu samples. These samples included representatives from various chondrite groups, including two splits of Orgueil (CI1), and one each of Tagish Lake (C2-ungrouped), Tarda (C2-ungrouped), Winchcombe (CM2), Murchison (CM2), Lancé (CO3.5), Vigarano (CV3), and Karoonda (CK4).

The elemental composition analysis showed that the 54 elements analyzed (Table 3, Figure 10A) in the Bennu aggregate samples from within TAGSAM closely align with the average composition of CI chondrites (Palme et al., 2014). CI chondrites have elemental abundances consistent with those of the solar photosphere, except the ice-forming elements (H, C, N, and O), Li, Be, B, and the noble gases (Palme et al., 2014). The Bennu material has higher abundances of the most volatile elements than the CM, CO, CV, CK, and ungrouped carbonaceous chondrites that were analyzed alongside it. Compared to samples returned from asteroid Ryugu, this Bennu sample exhibits a similar elemental composition, albeit without the refractory element enrichments (Nakamura et al., 2022; Yokoyama et al., 2023).

Like the TAGSAM sample, most elements in the avionics deck sample exhibit abundances similar to those found in CI chondrites. However, 10 elements are notably enriched: Na, P, Sb, Zr, Ba, La, Ce, Hf, Th, and U (Figure 10B). At the low end of these enrichments, P is enriched by 48% relative to CI chondrites. At the high end, U is enriched by a factor of 4. We ruled out the fibers as a source of these elements; SEM/energy-dispersive X-ray spectroscopy (EDS) data showed some Na but none of the other elements.

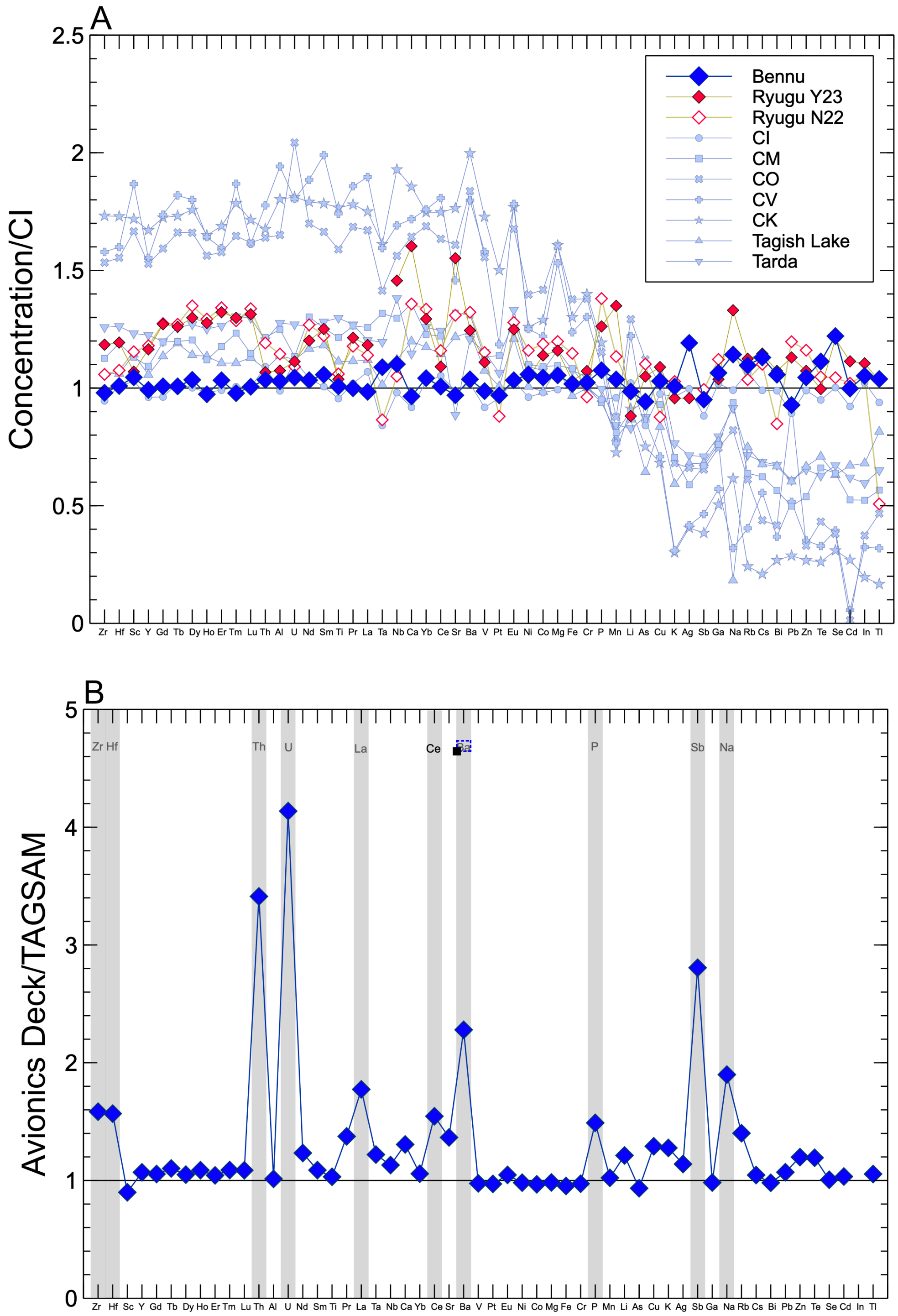

**Figure 10: Bulk concentrations of elements in order of increasing volatility.** (A) Concentrations of elements in Bennu material retrieved from inside TAGSAM compared to those measured for other astromaterials. Concentrations are normalized to the literature average value for CI chondrites as tabulated in Palme et al. (2014). (B) Concentrations of elements in a QL sample from the avionics deck, normalized to those in the sample shown above from inside TAGSAM, demonstrating elemental enrichments in the former.

**Table 3: Bulk elemental abundances in samples of Bennu and other carbonaceous astromaterials.**

| | Bennu (OREX-803015-0) | Bennu (OREX-501043-0) | Orgueil (CI1) | Orgueil (CI1) | Tagish Lake (C ung.) | Tarda (C ung.) | Winchcombe (CM2) | Murchison (CM2) | Lance (CO3.5) | Vigarano (CV3) | Karoonda (CK4) |
|---|---|---|---|---|---|---|---|---|---|---|---|
| Sample mass (mg) | 20.7 | 13.0 | 61.3 | 67.4 | 56.8 | 48.4 | 68.7 | 57.2 | 325.2 | 282.3 | 267.9 |
| Elemental abundances (parts per million by weight) | | | | | | | | | | | |
| Li | 1.49 | 1.80 | 1.54 | 1.34 | 1.31 | 1.25 | 1.60 | 1.36 | 1.95 | 1.84 | 1.38 |
| Na | 5826 | 11066 | 5054 | 6306 | 932 | 4664 | 3902 | 4785 | 4185 | 1630 | 3140 |
| Mg | 100303 | 98639 | 94534 | 93289 | 103241 | 110548 | 106692 | 104280 | 152691 | 145654 | 152890 |
| Al | 8614 | 8732 | 8271 | 7937 | 9523 | 10682 | 10475 | 10455 | 13814 | 16253 | 15089 |
| P | 1052 | 1567 | 930 | 913 | 932 | 974 | 906 | 917 | 1127 | 1065 | 1167 |
| K | 542 | 693 | 519 | 664 | 319 | 412 | 322 | 380 | 367 | 167 | 162 |
| Ca | 8527 | 11137 | 8114 | 8994 | 10594 | 10130 | 8392 | 12019 | 14525 | 15185 | 16404 |
| Sc | 6.1 | 5.5 | 6.0 | 5.6 | 6.7 | 7.2 | 6.8 | 6.6 | 9.7 | 10.9 | 10.0 |
| Ti | 453 | 467 | 430 | 424 | 501 | 584 | 565 | 547 | 715 | 784 | 795 |
| V | 53 | 52 | 49 | 49 | 58 | 62 | 63 | 61 | 83 | 85 | 93 |
| Cr | 2671 | 2598 | 2624 | 2525 | 2609 | 2729 | 2605 | 2611 | 3600 | 3399 | 3652 |
| Mn | 1965 | 2010 | 1818 | 1849 | 1492 | 1600 | 1582 | 1537 | 1663 | 1450 | 1376 |
| Fe | 188831 | 180036 | 187468 | 181755 | 179375 | 199608 | 198740 | 200952 | 255586 | 229607 | 241491 |
| Co | 531 | 513 | 498 | 485 | 500 | 546 | 535 | 554 | 720 | 619 | 655 |
| Ni | 11588 | 11375 | 10525 | 10529 | 11015 | 11952 | 11574 | 12043 | 15299 | 13698 | 13796 |
| Cu | 134 | 173 | 127 | 160 | 109 | 140 | 122 | 121 | 132 | 92 | 89 |
| Zn | 325 | 390 | 308 | 296 | 207 | 203 | 172 | 167 | 103 | 109 | 83 |
| Ga | 10.1 | 9.88 | 9.63 | 9.29 | 7.30 | 7.50 | 7.57 | 7.19 | 7.04 | 5.39 | 4.77 |
| As | 1.67 | 1.56 | 1.49 | 1.49 | 1.14 | 1.55 | 1.61 | 1.68 | 1.98 | 1.53 | 1.33 |
| Se | 24.9 | 25.1 | 20.4 | 19.2 | 12.9 | 13.7 | 11.8 | 13.0 | 7.76 | 8.04 | 6.32 |
| Rb | 2.43 | 3.41 | 2.37 | 2.08 | 1.66 | 1.59 | 1.32 | 1.42 | 1.36 | 0.90 | 0.54 |
| Sr | 7.55 | 10.3 | 7.70 | 7.88 | 9.48 | 6.90 | 7.74 | 10.3 | 12.5 | 11.4 | 13.7 |
| Y | 1.49 | 1.59 | 1.44 | 1.45 | 1.58 | 1.84 | 1.72 | 1.64 | 2.29 | 2.33 | 2.51 |
| Zr | 3.71 | 5.88 | 3.58 | 3.44 | 4.02 | 4.77 | 4.56 | 4.27 | 5.81 | 5.98 | 6.56 |
| Nb | 0.31 | 0.35 | 0.27 | 0.30 | 0.31 | 0.39 | 0.34 | 0.36 | 0.44 | 0.47 | 0.54 |
| Ag | 0.24 | 0.28 | 0.20 | 0.23 | 0.14 | 0.15 | 0.12 | 0.12 | 0.13 | 0.08 | 0.08 |
| Cd | 0.68 | 0.70 | 0.63 | 0.59 | 0.46 | 0.42 | 0.34 | 0.36 | 0.01 | 0.04 | 0.18 |
| In | 0.08 | 3.64 | 0.08 | 0.07 | 0.05 | 0.05 | 0.04 | 0.04 | 0.03 | 0.03 | 0.02 |
| Sb | 0.16 | 0.45 | 0.15 | 0.15 | 0.12 | 0.12 | 0.12 | 0.11 | 0.11 | 0.08 | 0.06 |
| Te | 2.57 | 3.07 | 2.19 | 1.95 | 1.64 | 1.45 | 1.19 | 1.53 | 1.00 | 0.76 | 0.60 |
| Cs | 0.21 | 0.22 | 0.19 | 0.17 | 0.13 | 0.13 | 0.11 | 0.12 | 0.08 | 0.10 | 0.04 |
| Ba | 2.48 | 5.65 | 2.43 | 2.45 | 2.96 | 2.89 | 2.53 | 3.13 | 4.39 | 4.30 | 4.77 |
| La | 0.24 | 0.43 | 0.26 | 0.23 | 0.27 | 0.29 | 0.29 | 0.31 | 0.41 | 0.46 | 0.43 |
| Ce | 0.63 | 0.98 | 0.66 | 0.61 | 0.72 | 0.78 | 0.75 | 0.77 | 1.02 | 1.13 | 1.10 |
| Pr | 0.09 | 0.13 | 0.10 | 0.09 | 0.11 | 0.12 | 0.11 | 0.12 | 0.16 | 0.18 | 0.17 |
| Nd | 0.49 | 0.60 | 0.49 | 0.47 | 0.55 | 0.60 | 0.57 | 0.61 | 0.80 | 0.89 | 0.85 |
| Sm | 0.16 | 0.18 | 0.15 | 0.14 | 0.18 | 0.20 | 0.19 | 0.19 | 0.25 | 0.30 | 0.27 |
| Eu | 0.06 | 0.06 | 0.06 | 0.06 | 0.07 | 0.08 | 0.07 | 0.07 | 0.10 | 0.10 | 0.10 |
| Gd | 0.21 | 0.22 | 0.20 | 0.20 | 0.24 | 0.25 | 0.24 | 0.25 | 0.33 | 0.36 | 0.36 |
| Tb | 0.04 | 0.04 | 0.04 | 0.04 | 0.05 | 0.05 | 0.05 | 0.05 | 0.06 | 0.07 | 0.07 |
| Dy | 0.26 | 0.27 | 0.25 | 0.25 | 0.29 | 0.32 | 0.30 | 0.30 | 0.42 | 0.45 | 0.44 |
| Ho | 0.05 | 0.06 | 0.06 | 0.06 | 0.06 | 0.07 | 0.07 | 0.06 | 0.09 | 0.09 | 0.09 |
| Er | 0.17 | 0.18 | 0.16 | 0.16 | 0.18 | 0.21 | 0.19 | 0.19 | 0.26 | 0.26 | 0.28 |
| Tm | 0.03 | 0.03 | 0.03 | 0.03 | 0.03 | 0.03 | 0.03 | 0.03 | 0.04 | 0.05 | 0.05 |
| Yb | 0.17 | 0.18 | 0.17 | 0.17 | 0.20 | 0.21 | 0.21 | 0.21 | 0.28 | 0.29 | 0.29 |
| Lu | 0.03 | 0.03 | 0.02 | 0.03 | 0.03 | 0.03 | 0.03 | 0.03 | 0.04 | 0.04 | 0.04 |
| Hf | 0.11 | 0.17 | 0.11 | 0.10 | 0.11 | 0.13 | 0.13 | 0.13 | 0.16 | 0.17 | 0.18 |
| Ta | 0.02 | 0.02 | 0.01 | 0.02 | 0.02 | 0.02 | 0.02 | 0.02 | 0.02 | 0.02 | 0.02 |
| Pt | 0.90 | 0.88 | 0.89 | 0.87 | 0.94 | 0.99 | 0.97 | 1.06 | 0.51 | 1.10 | 1.40 |
| Tl | 0.15 | 0.15 | 0.13 | 0.12 | 0.11 | 0.09 | 0.08 | 0.08 | 0.07 | 0.04 | 0.02 |
| Pb | 2.45 | 2.62 | 2.36 | 2.38 | 1.59 | 1.61 | 1.25 | 1.31 | 2.46 | 1.36 | 0.76 |
| Bi | 0.12 | 0.12 | 0.11 | 0.11 | 0.08 | 0.08 | 0.06 | 0.06 | 0.05 | 0.04 | 0.03 |
| Th | 0.03 | 0.11 | 0.03 | 0.03 | 0.03 | 0.04 | 0.04 | 0.04 | 0.05 | 0.05 | 0.05 |
| U | 0.01 | 0.04 | 0.01 | 0.01 | 0.01 | 0.01 | 0.01 | 0.01 | 0.02 | 0.01 | 0.01 |

### *Hydrogen, Carbon, and Nitrogen Abundances and Isotopic Compositions*

Elemental and isotopic analyses of bulk hydrogen, carbon, and nitrogen were carried out at the Carnegie Institution for Science using a Thermo Scientific Delta Vplus isotope ratio mass spectrometer (IRMS) coupled with a Carlo Erba elemental analyzer (EA) for C and N analyses and a Thermo Scientific Delta Q IRMS connected to a Thermo Finnigan Thermal Conversion EA for H analyses. As with the bulk elemental abundances, we analyzed one QL sample from the avionics deck and one sample from inside TAGSAM (Table 1). The samples consisted of a mixture of fine- to intermediate-sized particles. We split them into subsamples of 1–1.5 mg and 5.5 mg for H and C/N analyses, respectively, and each measurement involved the analysis of at least three replicate subsamples. For one split from the QL aggregate, a subsample of intermediates (~200 μm) was hand-picked from the fines and analyzed separately. For all other splits, subsampling was performed without any further particle size fractionation.

Samples underwent heating at 120°C for 48 hours in an argon-flushed glovebox to remove adsorbed water. Prior to the H-C-N analysis, the samples were re-weighed and transferred to the EA-IRMS system without exposure to air.

The two samples analyzed have the same bulk hydrogen, carbon, and nitrogen abundances within the measurement uncertainties (Table 4). Further, the abundances in the fines and intermediates in the TAGSAM sample are identical within the margin of error.

The total average bulk carbon content ranges between 4.5 and 4.7 wt.%. This abundance is higher than that of any meteorite analyzed at the Carnegie Institution for Science and the published bulk carbon data for Ryugu (Naraoka et al., 2023). The isotopic ratios of hydrogen and nitrogen are distinct from those of other carbonaceous astromaterials (Figure 11A). The C isotopic ratios combined with C/N ratio suggest that about 10% of the total carbon in the samples may be in the form of carbonates, with the remaining 90% in organic molecules.

### *Bulk Oxygen Isotopes*

Bulk O isotopic composition was determined using a laser-assisted fluorination system at the Open University (Table 4). Two samples were analyzed, consisting of approximately 5 mg of fines and 3 mg of intermediate particles from the QL allocation (Table 1). These samples were subsequently split into replicates. In this case, the QL samples were not protected from air exposure after their removal from JSC's curation facility.

A modified “single shot” method was employed, like that used for Ryugu samples (Greenwood et al., 2023), which involved heating the samples with a 10.6-μm $CO_2$ laser in the presence of $BrF_5$ to liberate $O_2$ gas. The released gas was then purified and analyzed using a Thermo MAT 253 dual-inlet mass spectrometer. The weighted average of the four analyses is 11.4 ± 1.6‰ for $\delta^{17}O$, 20.9 ± 2.7‰ for $\delta^{18}O$, and 0.75 ± 0.17‰ for $\Delta^{17}O$. The variation displayed by the samples exceeds typical analytical precision (±0.05‰ for $\delta^{17}O$, ±0.10‰ for $\delta^{18}O$, and ±0.02‰ for $\Delta^{17}O$). A small blank correction was applied to the results, accounting for less than 3% of the smallest sample. Figure 11B shows the results compared with data from other carbonaceous astromaterials.

**Table 4: H, C, N, O abundances and isotopic ratios in Bennu samples.**

| Sample | Type | H (wt%) | $\delta$D (‰) | C (wt%) | $\delta^{13}$C (‰) | N (wt%) | $\delta^{15}$N (‰) |
|---|---|---|---|---|---|---|---|
| OREX-500002-0 | Fines | 0.90 ± 0.04 | 315 ± 2 | 4.7 ± 0.4 | 3.2 ± 0.1 | 0.23 ± 0.02 | 75.5 ± 0.1 |
| OREX-500002-0 | Intermediates | 0.93 ± 0.05 | 305 ± 2 | 4.7 ± 0.4 | -0.5 ± 0.1 | 0.24 ± 0.02 | 57.1 ± 0.1 |
| OREX-803007-0 | Aggregate | 0.93 ± 0.01 | 344 ± 18 | 4.5 ± 0.2 | 3.3 ± 1.3 | 0.25 ± 0.01 | 82 ± 21 |
| | | **$\delta^{17}$O (‰)** | **$\delta^{18}$O (‰)** | **$\Delta^{17}$O (‰)** | | | |
| OREX-501066-0 | Fines | 11.41 ± 0.08 | 20.59 ± 0.02 | 0.70 ± 0.09 | | | |
| OREX-501042-0 | Fines | 12.39 ± 0.07 | 22.22 ± 0.02 | 0.84 ± 0.08 | | | |
| OREX-501067-0 | Intermediates | 11.40 ± 0.07 | 20.52 ± 0.17 | 0.73 ± 0.02 | | | |
| OREX-501047-0 | Intermediates | 10.45 ± 0.02 | 18.87 ± 0.02 | 0.63 ± 0.03 | | | |

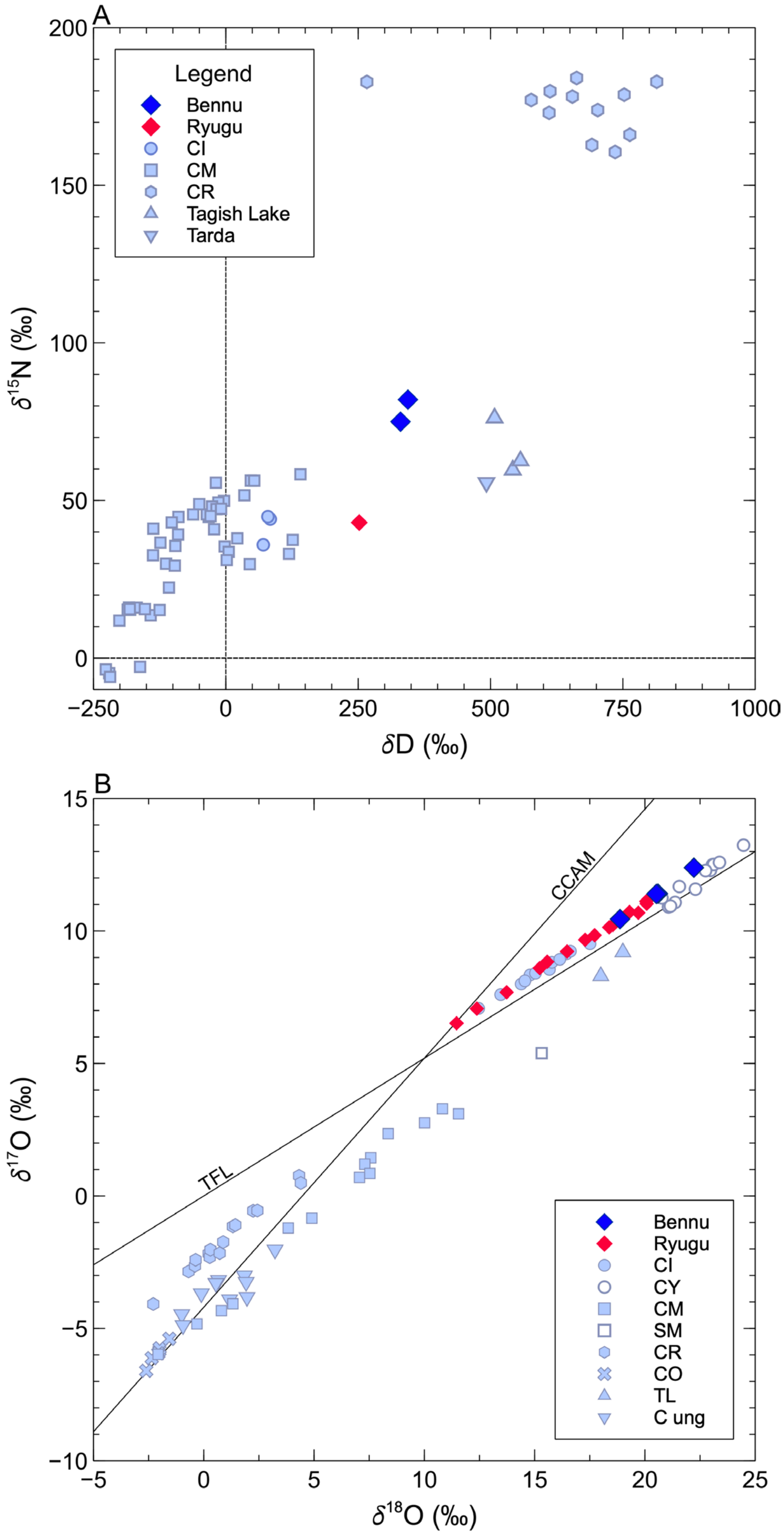

**Figure 11: H, N, and O isotopic systematics.** (A) Hydrogen and nitrogen isotopic of analyzed Bennu samples compared to other carbonaceous astromaterials. The dashed vertical and horizontal lines indicate the average composition of terrestrial materials. (B) As above for oxygen isotopes. TFL stands for the terrestrial fractional line and CCAM is the carbonaceous chondrite anhydrous mixing line. H,C isotope data for meteorites from Alexander et al. (2012, 2013, 2015, 2018), Davidson et al. (2014, 2019), Nittler et al. (2021),

and Bischoff et al. (2021). O isotope data from Greenwood et al. (2023), Yokoyama et al. (2022), Nakamura et al. (2022), Clayton and Mayeda (1999), Lee et al. (2016), Hewins et al. (2014), Kimura et al. (2020), Schrader et al. (2011), Alexander et al. (2018), Jacquet et al. (2016), Ruzicka et al. (2015), Gattacceaca et al. (2020), Brown et al. (2000).

## Mineralogical and Chemical Analysis of Individual Phases

### *Mineral Volumetric Abundances*

A QL sample, which consisted of >80 mg of fines collected from the avionics deck (Table 1), was analyzed using powder X-ray diffraction (PXRD) facilities at JSC. Fine particles were gently pressed onto a low-background front-loading sample holder to achieve a flat surface for precise measurements, minimizing sample handling and preparation. No grinding was performed on the sample to avoid introducing friction that could cause phase changes. The sample was analyzed using a Malvern Panalytical X'Pert Pro PXRD in "Reflection-Transmission" configuration with data collected every 0.006° at 40 mA and 45 kV from 2° to 90° (2θ) on a spinner stage in ambient conditions. Patterns were acquired using Co Kα radiation (λ = 1.78901 Å) with an Fe Kβ filter used to eliminate fluorescence.

Mineral identification and quantitative phase analysis were performed using the X'Pert HighScore software package. A Rietveld approach was used for an assessment of the modal mineralogy with 28 variables including one profile parameter for all phases except magnetite, which had three parameters but no preferred orientation parameters. The final $R_p$ and $R_{wp}$ values were 3.6% and 5%, respectively. $R_p$ (Rietveld profile factor) is a measure of the agreement between the observed and calculated diffraction patterns. $R_p$ values closer to zero indicate better agreement between the observed and calculated patterns. $R_{wp}$ (weighted profile factor) is another measure of the agreement between the observed and calculated diffraction patterns, like $R_p$. However, $R_{wp}$ is weighted by the square root of the observed intensities, giving more weight to the higher-intensity peaks.

The PXRD results (Figure 12) show that phyllosilicates are the dominant mineral phase, constituting approximately 80% of the volume. Sulfides account for about 10% of the volume, while magnetite, carbonate, and olivine contribute around 5%, 3%, and 2%, respectively.

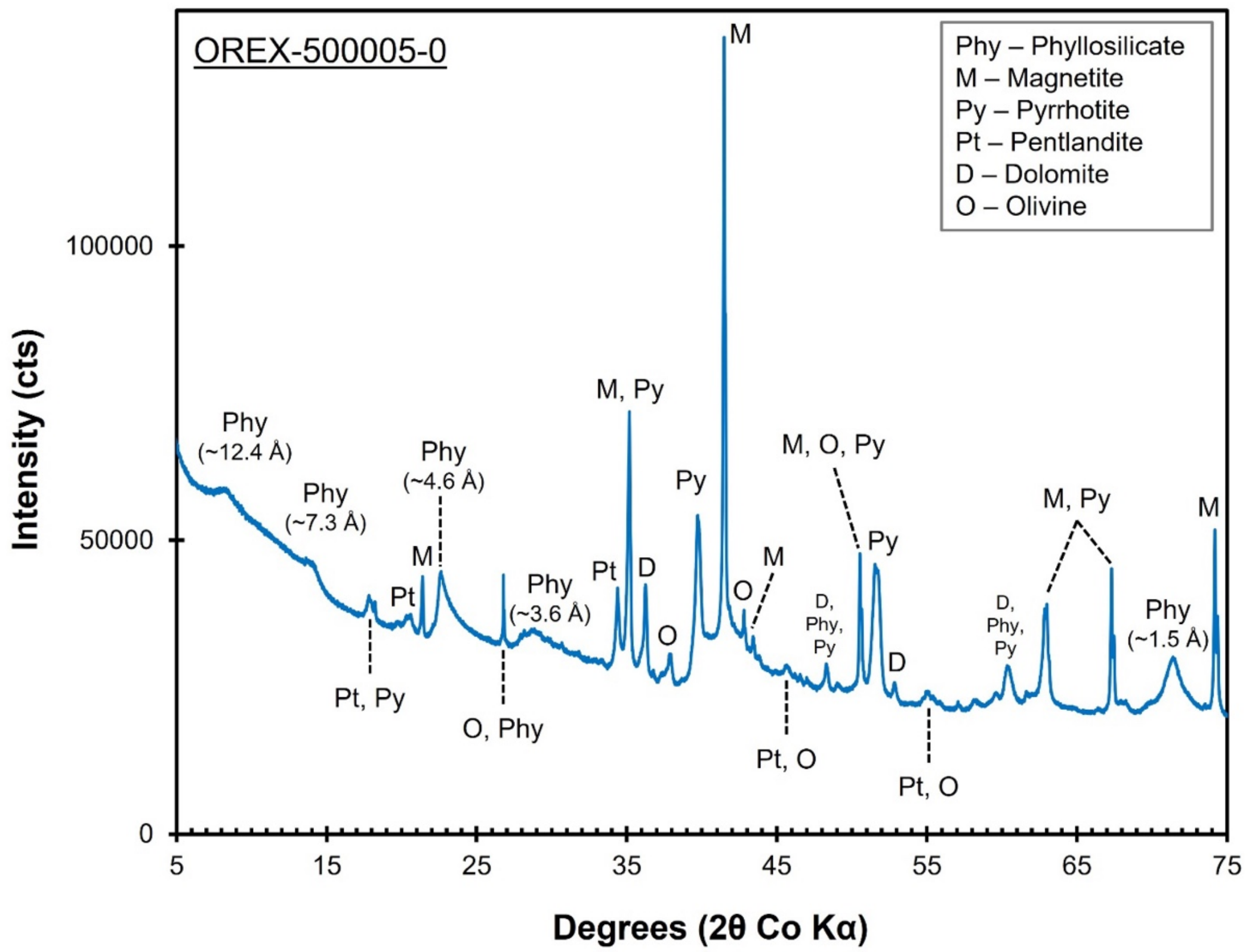

**Figure 12: XRD spectra and mineral identification.**

### *Mineral Identification and Associations*

Three grain mounts were prepared by sub-sampling three of the QL aggregate samples: one with fines and two with intermediate particles (Table 1). The mounts were carbon-coated before analysis. We used JEOL 7600F and 7900F SEMs equipped with Oxford Instruments Ultim Max EDS detectors at JSC to capture high-resolution images and EDS spectra (Figures 13–16) and a Hitachi TM400Plus low-vacuum tabletop SEM at the University of Arizona (Figure 17).

From these observations, the prevailing component is Mg-bearing phyllosilicates, primarily serpentine and smectite (Figure 13). The phyllosilicates vary in particle size from fine (sub-micron) to coarse (hundreds of microns). The coarse phyllosilicates occur in distinct pods with a rosette-type structure and exhibit fewer associated minerals (Figure 13A). In contrast, the fine phyllosilicates are associated with other minerals, dominantly Fe-sulfides, magnetite, and carbonates (Figure 13B).

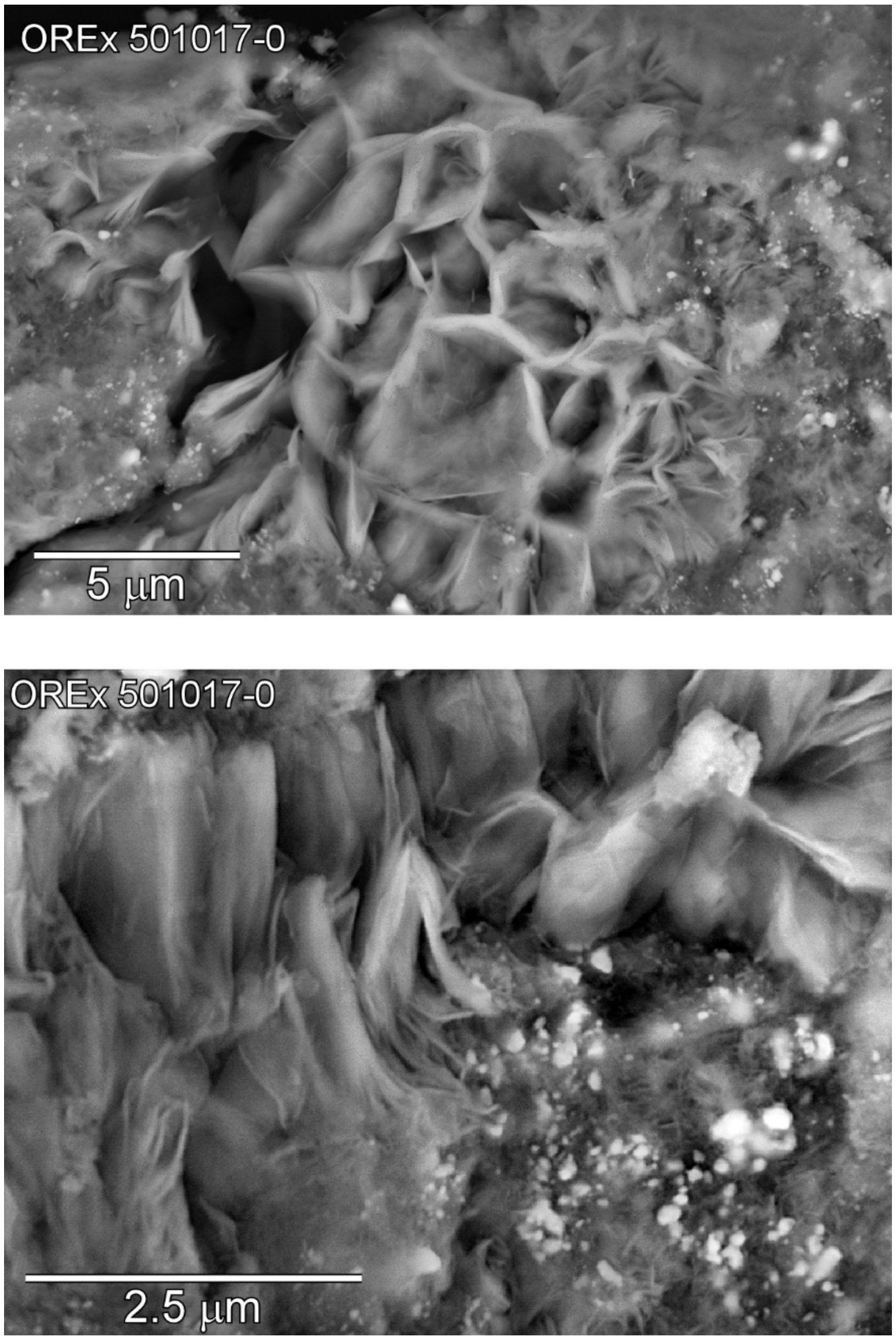

**Figure 13: Examples of coarse-grained (A) and fine-grained (B) phyllosilicates.**

Magnetite is plentiful in both isolated grains and framboids, often concentrated in distinct regions and veins. These magnetite grains exhibit diverse morphologies, including framboids (both discrete and in aggregates; Figure 14A), spheres with radially projecting needles (Figure 14B), dodecahedral forms (Figure 14B,C), and plaquettes (Figure 14D). Sulfides, primarily pyrrhotite with lesser quantities of pentlandite and other trace phases, are also present. The abundance and habit of sulfides vary, including large pyrrhotites with euhedral shapes or visible cross-sections, as well as fine dispersed anhedral particles. The majority of pyrrhotite grains appear as pseudohexagonal plates (Figure 14E), sometimes forming stacks or blocky aggregates (Figure 14F). Both sulfides and magnetite exhibit surface etching features, appearing as irregular pits or cavities, often filled with secondary minerals (Figure 14C).

Carbonates, including dolomite, calcite, magnesite, and Mg- and Mn-bearing breunnerite, are dispersed throughout the sample as individual grains (Figure 15A) or concentrated assemblages (Figure 15B,C). They are often found in association with magnetite and pyrrhotites. Surface alteration features, similar to those observed on the oxides and sulfides, are also present in the carbonates, although they are less pronounced (Figure 15D).

Within the mineral matrix, various carbonaceous phases are interspersed. Discrete plates and/or veins of carbonaceous material are typically thin, sheet-like structures or narrow, elongated features embedded within the phyllosilicate matrix. In SEM images, these structures appear as dark, carbon-rich areas contrasting with the surrounding minerals (Figure 16A). Aggregates with carbonaceous minerals in are composed of various mineral and organic components that are fused or bound together (Figure 16B). Organic nanoglobules have a spherical structure, ranging in size from nanometers to micrometers (Figure 16C). These structures often contain a dense, solid core surrounded by a more diffuse organic shell. In other cases, the cores may be hollow or contain voids.

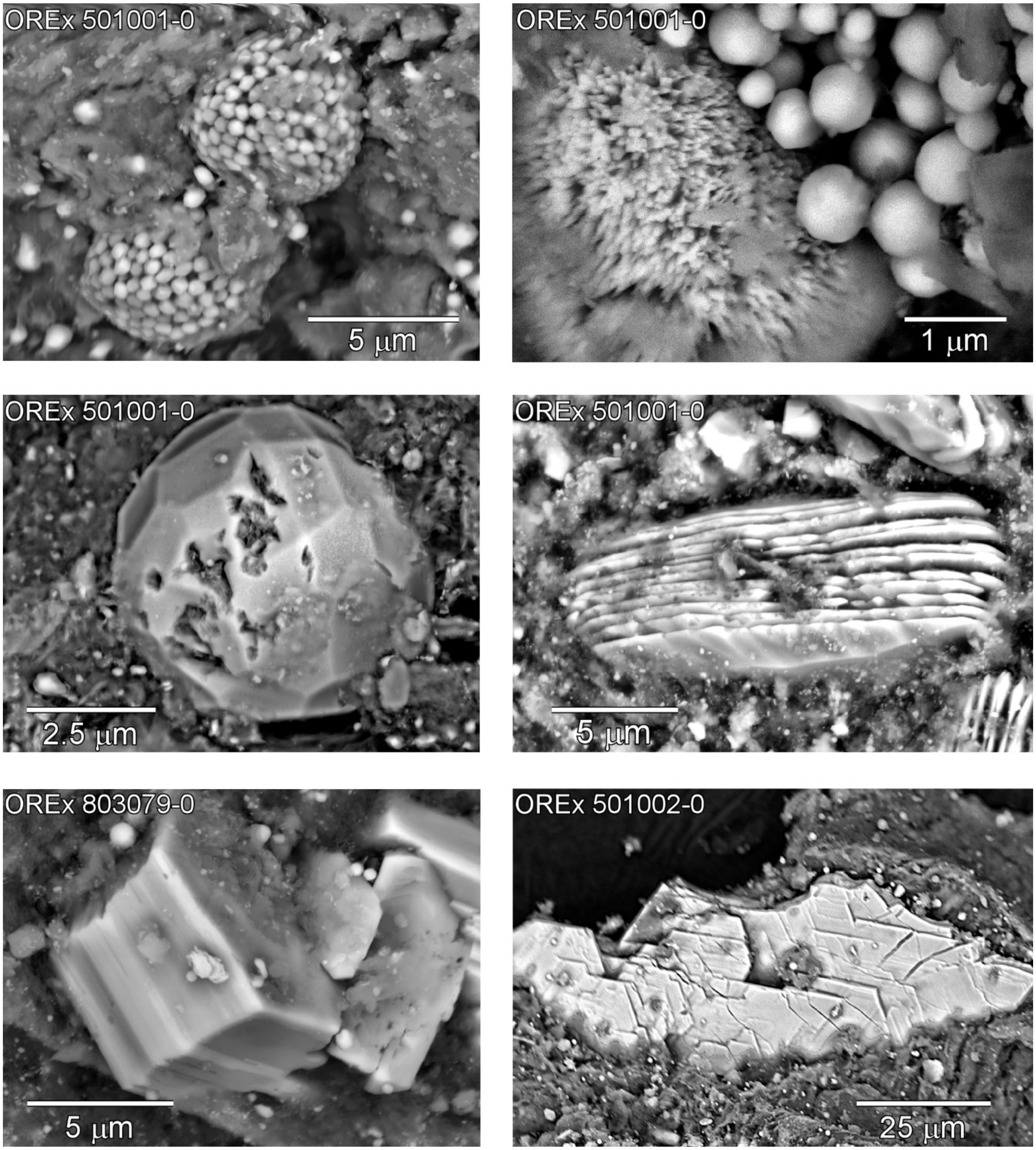

**Figure 14: Examples of magnetite and sulfide.** (A) Magnetite framboids. (B) Magnetite spheres with radially projecting needles next to dodecahedral crystals. (C) Dodecahedral magnetite with surface etching. (D) Magnetite plaquettes. (E,F) Pseudohexagonal plates of pyrrhotites.

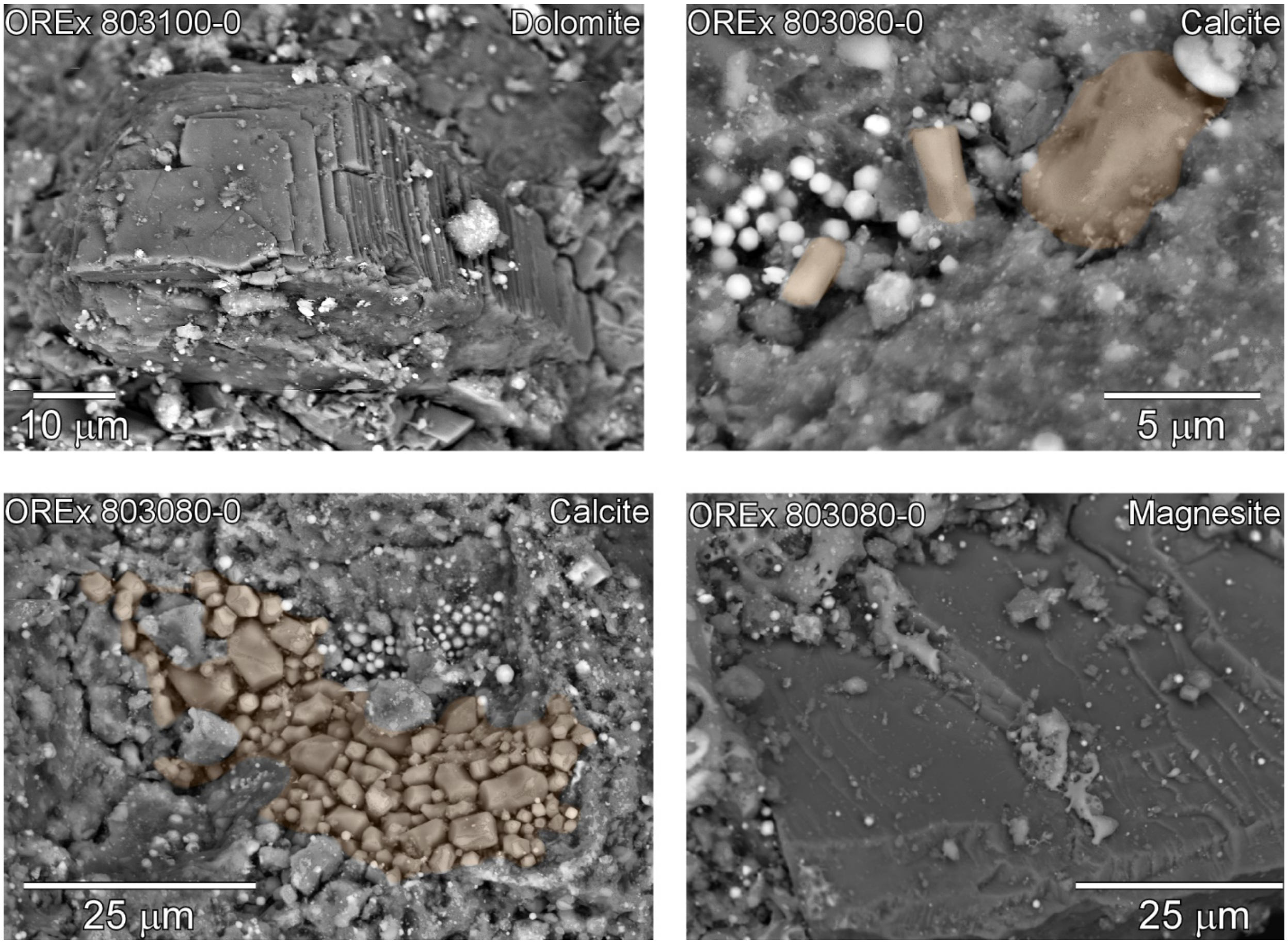

**Figure 15: Examples of carbonates.** (A) Individual dolomite grain. (B,C) Calcite assemblages. (D) Magnesite with surface alteration features.

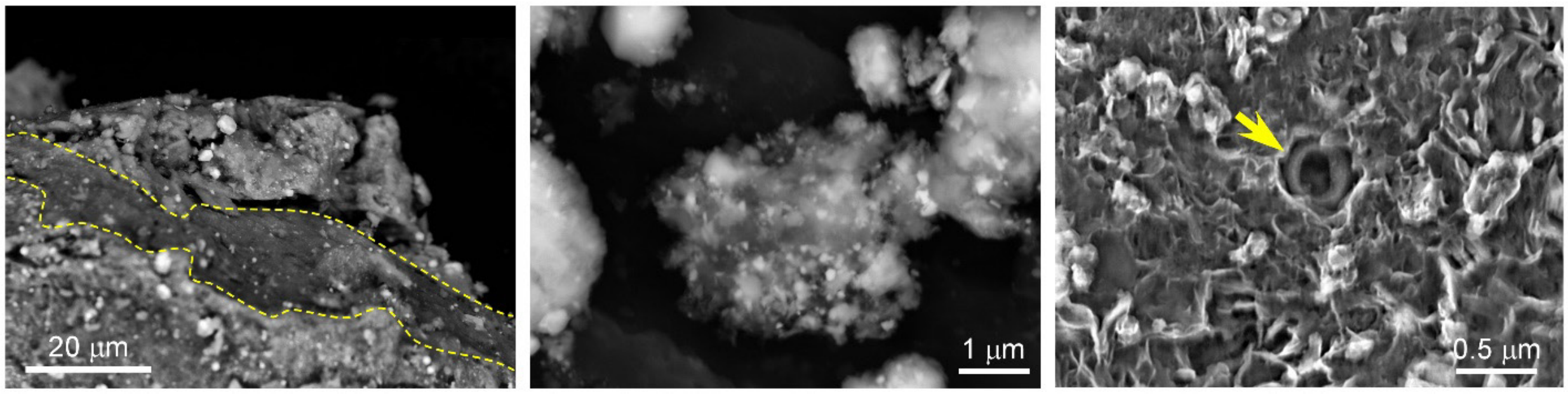

**Figure 16: Examples of the different types of condensed organic phases and minerals.** (A) Carbon-rich area distinct from surrounding minerals (outlined in yellow). (B) Carbon–mineral aggregate. (C) Organic nanoglobule (yellow arrow).

Minor and trace phases identified include Mg,Na-phosphate, Ca-phosphate, Fe,Ni-phosphide (potentially schreibersite), Fe,Cr-phosphide (possibly andreyivanovite), Mn-bearing chromite, Mn-bearing ilmenite, and dendritic Cu-sulfide. Some particle surfaces contain vesiculated melt droplets, indicative of space weathering processes.

Phosphates are present in various forms, including isolated particles, euhedral to anhedral grains, and as veins and surface coatings. Many of the isolated bright particles within the aggregates are composed of Mg,Na-phosphate. To gain insight into the petrographic context of this material, we examined a millimeter-sized mottled particle. In optical, reflected light images, this particle displays a distinct high-reflectance phase covering a dark groundmass (Figure 17A). SEM images reveal bright hexagonal pyrrhotite and framboidal magnetite dispersed throughout the darker material, with small veins of high-reflectance material present within the groundmass (Figure 17B). During transfer to an SEM mount, the particle fragmented along one of these veins, revealing material similar in appearance to the outer crust (Figure 17C). SEM characterization indicates that the phase is tens of microns in thickness, occurring as blocky friable fragments with a texture suggestive of desiccation (Figure 17D). EDS analysis identified this material as primarily composed of Na, Mg, and P.

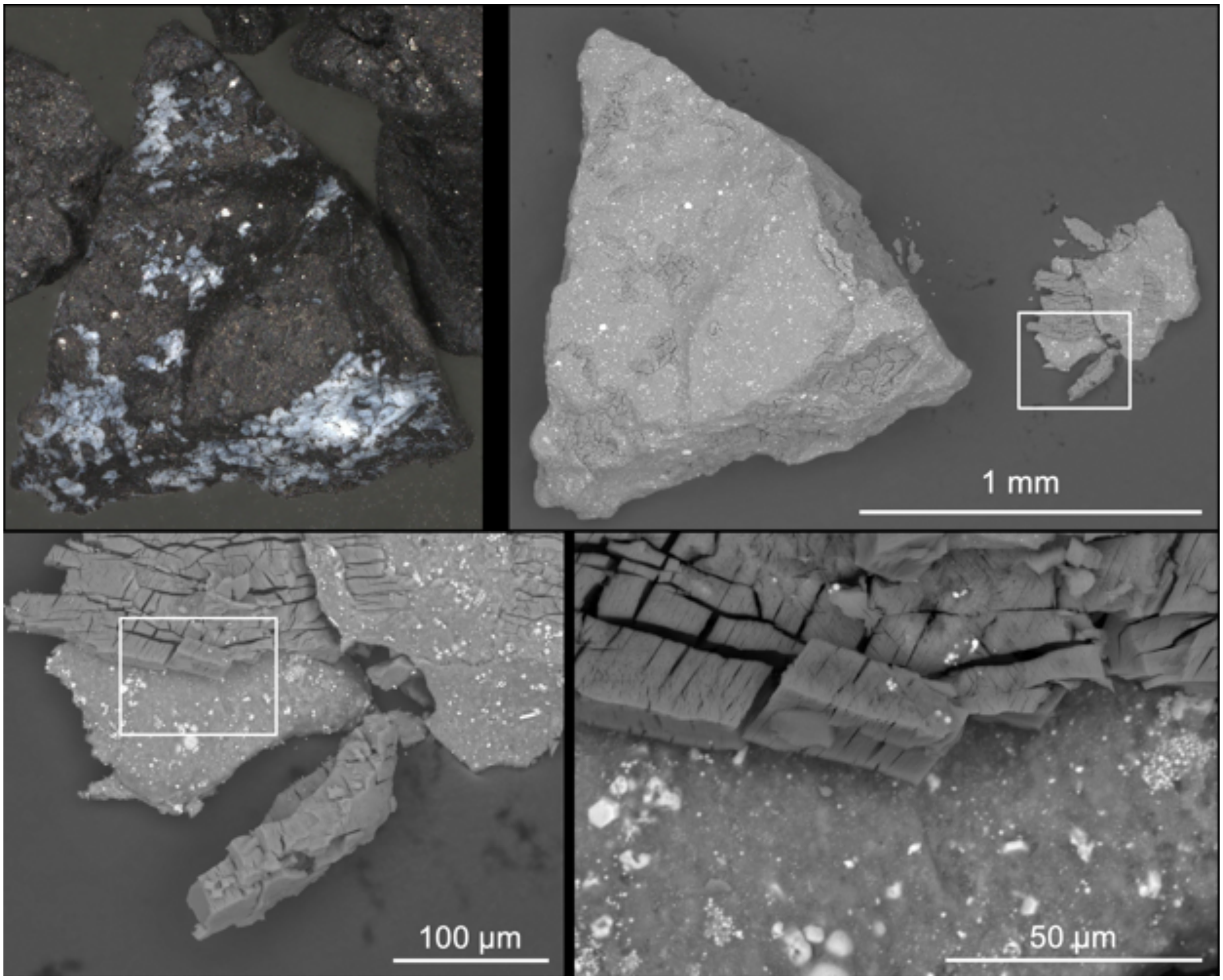

**Figure 17: Phosphate in a mottled particle.** (A) Visible light microscopy image of a dark particle with an outer crust of high-reflectance material. (B–D) SEM images showing progressively zoomed view of a fragment of the particle that split off along a high-reflectance vein, revealing material similar to the outer crust, with a blocky friable texture and consisting of Na, Mg, and P.

### *Mineral Point Counting Statistics*

Approximately 2 mg of material from a QL aggregate (Table 1) were distributed onto a concavity glass slide for closer examination of the modal mineralogy using SEM combined with EDS. The fines in this subsample tended to agglomerate, forming clusters up to ~0.5 mm. Isolated bright white particles were a common minor constituent. One highly reflective particle composed of pure aluminum was likely a spacecraft contaminant.

This material was further subsampled and segregated into fine, intermediate, and coarse size fractions. The fines were transferred to beryllium planchettes, and the intermediate particles were affixed to aluminum cylinder mounts using carbon tape. Subsequently, a thin sputter-coating of carbon was applied to mitigate sample charging effects.

We used the JSC JEOL 7600F and 7900F SEMs to acquire EDS point spectra and generate detailed elemental maps. The SEMs operated with a 15 kV electron beam at an approximate current of 1 nA. AZtec “Point & ID” and “Feature” software packages were used for data reduction and analysis. A comprehensive set of EDS spectra with mineral identifications was collected from a total of 635 individual particles. The elemental abundances were used to classify each point into one of several categories (Figure 18).

The results, in descending order of abundance, show that phyllosilicates — specifically, mixed serpentine- and smectite-group minerals — constitute 56.7% of the measured particles. A significant portion of the particles, 13.2%, is composed of Fe-rich material with varying sulfur and nickel content. In addition, 2.1% of the measured phases are identified as pyrrhotite, characterized by stacked hexagonal plates. Pentlandite is a trace component, representing about 0.3%. Combined, 15.6% of this sample consists of Fe,Ni-sulfide phases.

Magnetite is identified in 9.6% of the measured points, displaying various morphologies, including weathered/etched surfaces, plaquettes, and framboids. Carbonates make up 3.6% of the particles. The range of cation contents in the carbonates indicate the presence of calcite, dolomite, and breunnerite. Anhydrous silicates are also present in small amounts, with forsteritic olivine and low-calcium pyroxene constituting 3.2% and 0.8% of the composition, respectively. Only 10.6% of the measured points are unclassified, indicating unidentified mineral phases or mixtures of materials.

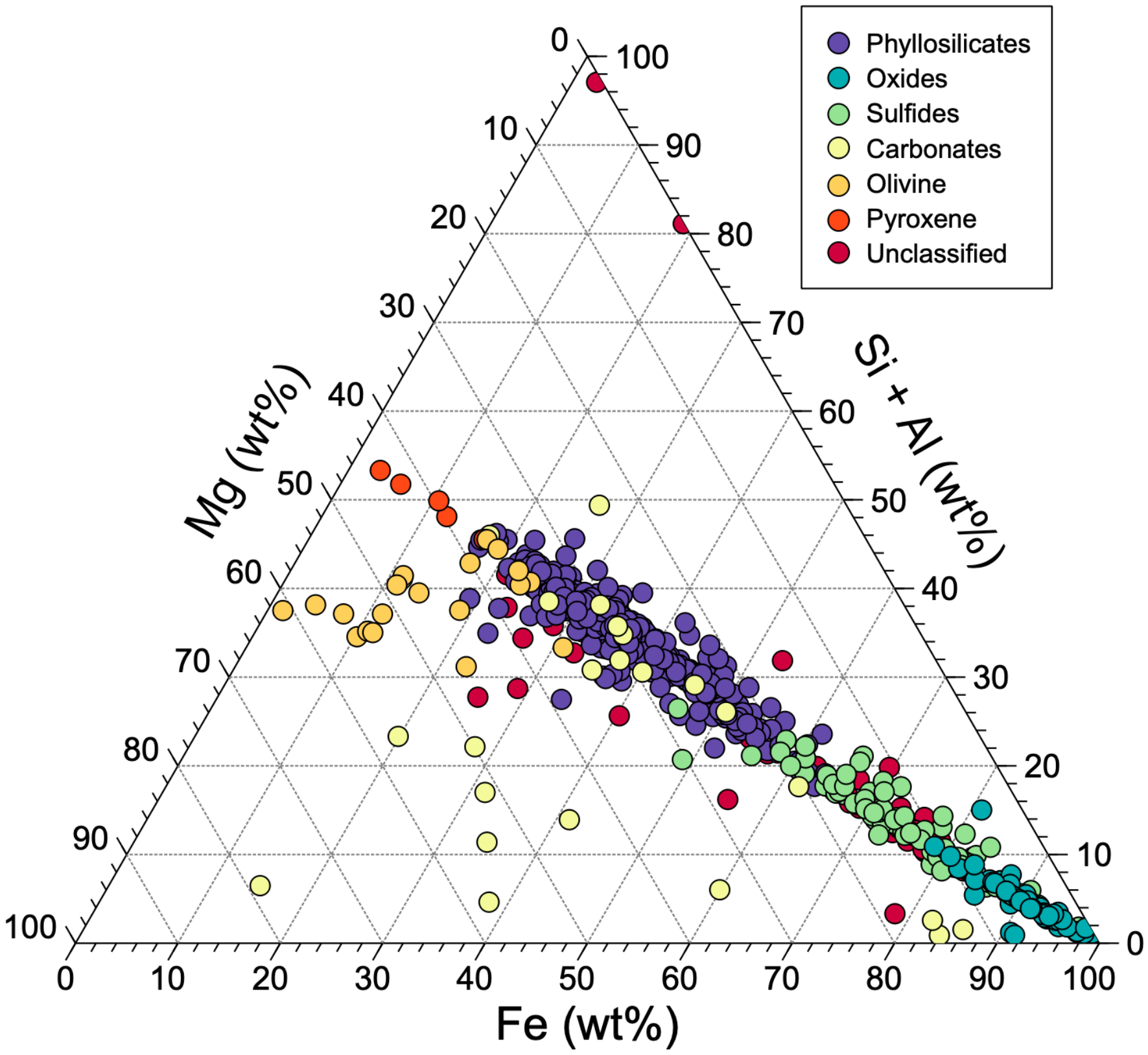

**Figure 18: Ternary diagram of major element abundances.** Data are normalized to Mg, Fe, and Si + Al and minerals are identified based on the calculated stoichiometries. Points labeled "unclassified" did not fit any known mineral species and are likely mixtures of materials.

### *Mineral Spectroscopy*

Micro-FTIR (μFTIR) was used to identify minerals and organic molecules in a subset of a QL aggregate sample (Table 1). A small batch of fine particles was pressed into a KBr disk and were analyzed with a Bruker Vertex 70 (bench) and a Hyperion 3000 (microscope) in transmission mode at JSC. The microscope's stage allowed for sample mapping from the micron to the millimeter scale, using the mercury cadmium telluride detector with a spectral range of 2.5–16.7 μm (4000–600 $cm^{-1}$).

Most of the μFTIR spectra show a distinct OH feature at 2.73 μm (3670 $cm^{-1}$) and the presence of the characteristic 10-μm absorption feature indicative of phyllosilicates (Figure 19A). Peaks corresponding to carbonates dominate many of the spectra, with clear identification of calcite and dolomite (Figure 19B,C). For example, the $CO_2$ bending mode occurs around 4.3 μm. Carbonate minerals, such as calcite and dolomite, exhibit strong absorption features in the mid-infrared region, including around 6.9 and 11.6 μm, due to the vibrational modes of carbonate ions. Dolomite typically shows a characteristic absorption feature around 14 μm, whereas calcite exhibits a strong absorption feature around 14.5 μm, both of which are related to the presence of carbonate ions in its structure. Additionally, weak aliphatic bands at around 5.6 µm related to C–H stretching vibrations are present in many of the spectra, characteristic of organic molecules. These data indicate that phyllosilicates are abundant in the samples and are intergrown with carbonates, organics, and other minor phases.

This subsample included many small bright particles, consistent with other sample splits. µFTIR analyses of these particles (Figure 19D) reveals prominent spectral absorptions at 6, 7, 9, and 11 µm. These spectra match those of magnesium hydrogen phosphate trihydrate ($MgHPO \cdot 3H_2O$) (Yu et al., 2016). This material is a white crystalline solid that is soluble in water (Verbeek et al., 1984).

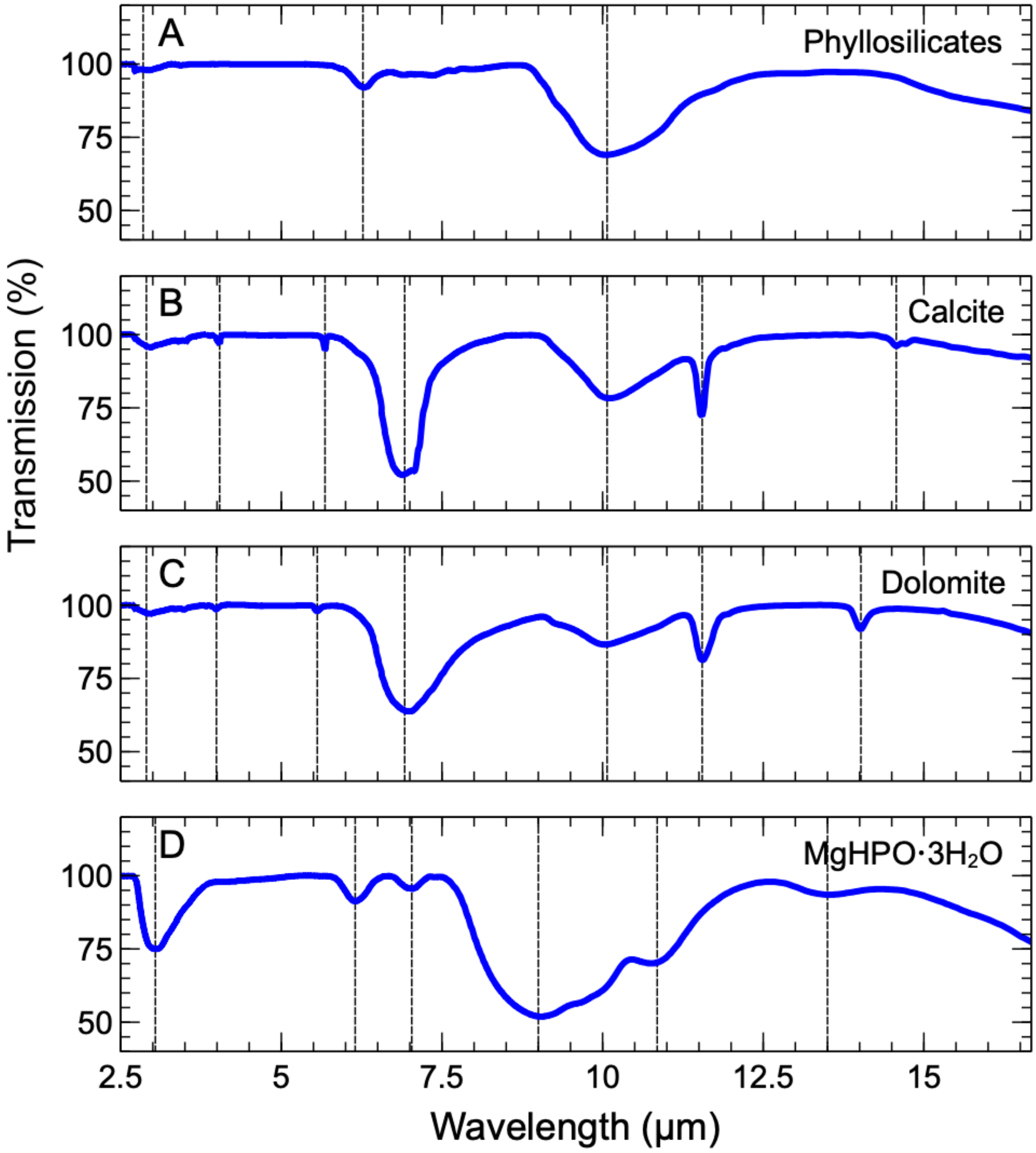

**Figure 19: µFTIR spectra.** Spectra obtained from individual mineral grains provide definitive identification of the phyllosilicates, carbonates, and phosphates.

### *Carbon Distribution*

A subset of the QL samples (Table 1) underwent characterization at JSC for carbon distribution. Initial documentation involved acquisition of large-area optical photomosaics. From these, regions of interest were identified and re-imaged at higher resolution, with extended depth of field, both optically and under ultraviolet (UV) fluorescence. For UV fluorescence, a broadband (330–385 nm) Hg-arc lamp was used for excitation in combination with a long-pass (>420 nm) emission filter. All high-resolution images were captured using a Nikon BX-60 microscope at a pixel resolution at least twice the Abbe diffraction limit (i.e., $\lambda/2NA$, where $\lambda$ is the wavelength of light used for imaging and NA is the numerical aperture of the microscope objective).

Spatially correlated organic analysis of these samples was then performed using microprobe two-step laser mass spectrometry (µL2MS). Point spectra and 2D maps were acquired at a spatial resolution of 5 µm using either a UV (266 nm) or vacuum-UV (VUV; 118 nm) photoionization source. The former allows for selective detection of polycyclic aromatic hydrocarbons (PAHs) by 1 + 1 resonance-enhanced multiphoton ionization, and the latter enables the broad detection of most organic species via non-resonant single photon ionization.

Blue fluorescence is spatially correlated with the presence of carbonate and phosphate minerals embedded in the phyllosilicate groundmass (Figure 20). The fine-grained matrix also exhibits discrete yellow fluorescence hotspots that have been demonstrated by SEM/EDS to be organic nanoglobules (Figure 16C). This intrinsic nanoglobule fluorescence was observed to be thermolabile, being quenched under extended electron beam exposure. This observation implies that fluorescent nanoglobules in the Bennu samples did not experience consequential heating after their formation.

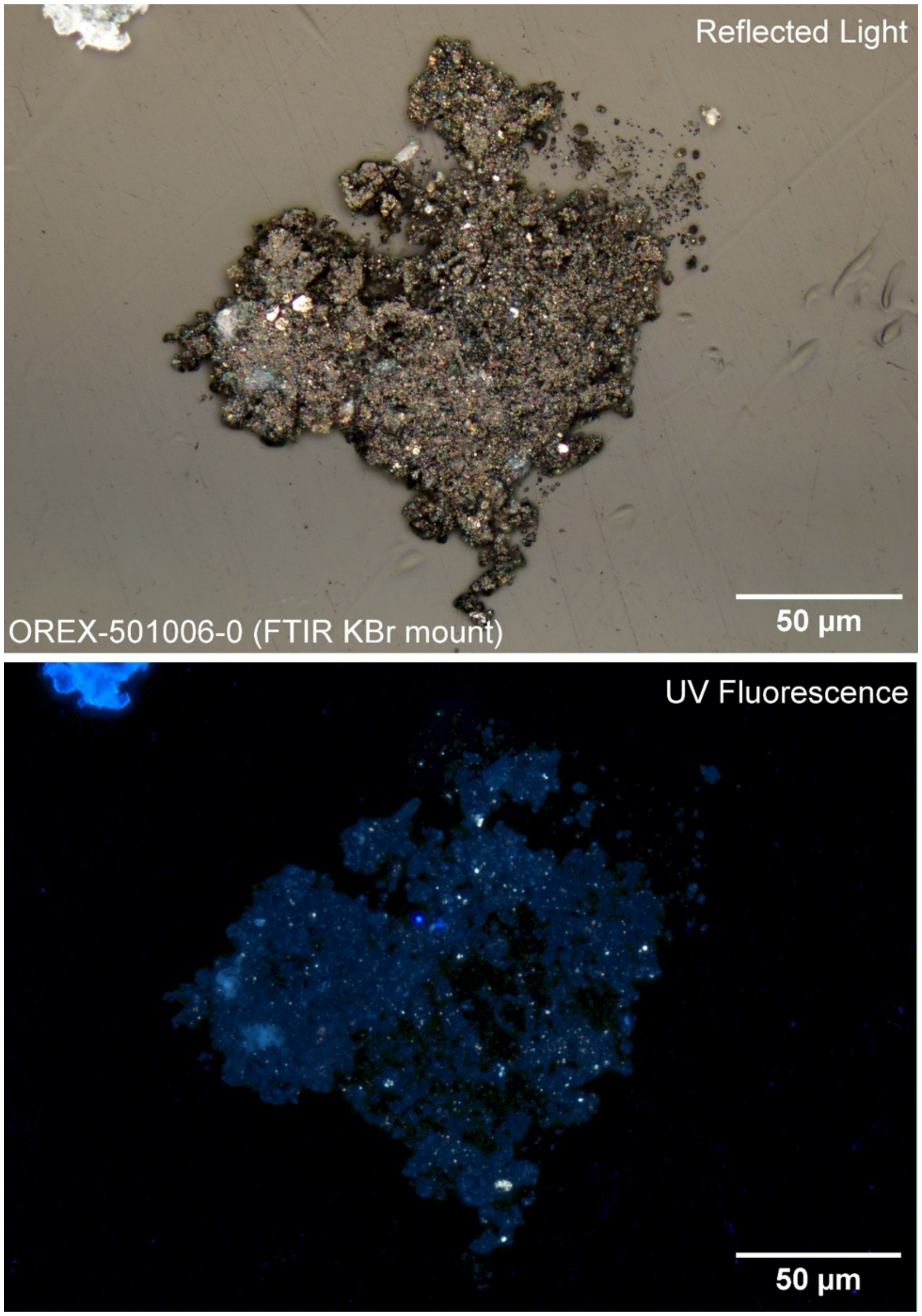

**Figure 20: Optical and UV fluorescence imaging.** (Top) A reflected light image showing the texture and brightness variation across the analyzed particle. (Bottom) UV fluorescence microscopy image showing the distribution of carbonates and phosphates (blue fluorescence) and organic nanoglobules (yellow fluorescence).

The µL2MS analysis involved the examination of a subsample of fine particles from a QL aggregate (Table 1) crushed onto an optical KBr platter. Upon loading the sample platter into the µL2MS main vacuum chamber, a considerable increase in chamber background pressure was observed and a mass spectrum was collected (Figure 21). In addition to the sample mounted on an optical KBr platter, we included a reference sample consisting of powdered Allende (CV3) meteorite on Al foil. Both the reference meteorite and sample platter had been degassed for about six months prior to loading. Thus, we attribute the elevation in pressure to sample volatile outgassing, rather than the sample platter or the Allende reference.

After the pressure returned to nominal operating conditions, we performed the µL2MS analyses on three distinct clusters of particles. We collected point spectra and 2D spatial mapping data using both the UV (266 nm) and VUV (118 nm) photoionization techniques. The particle spectra closely match the initial degassing spectrum (Figure 21). These particles are rich in organics, primarily composed of 1- to 4-ring PAHs. Lower-mass PAHs are less abundant than expected based on comparison to carbonaceous chondrites, and there are no apparent signs of significant thermal processing, except for a 254 m/z peak that warrants further investigation. One particle cluster contains a heterogeneous distribution of PAHs within the matrix, with individual PAH species exhibiting spatial variation that correlates with the fluorescent hotspots. There is no evidence of terrestrial contamination in the spectra. In comparison to other known astromaterials, these Bennu sample spectra most closely resemble those of the Orgueil (CI1) carbonaceous chondrite (Clemett et al., 1998).

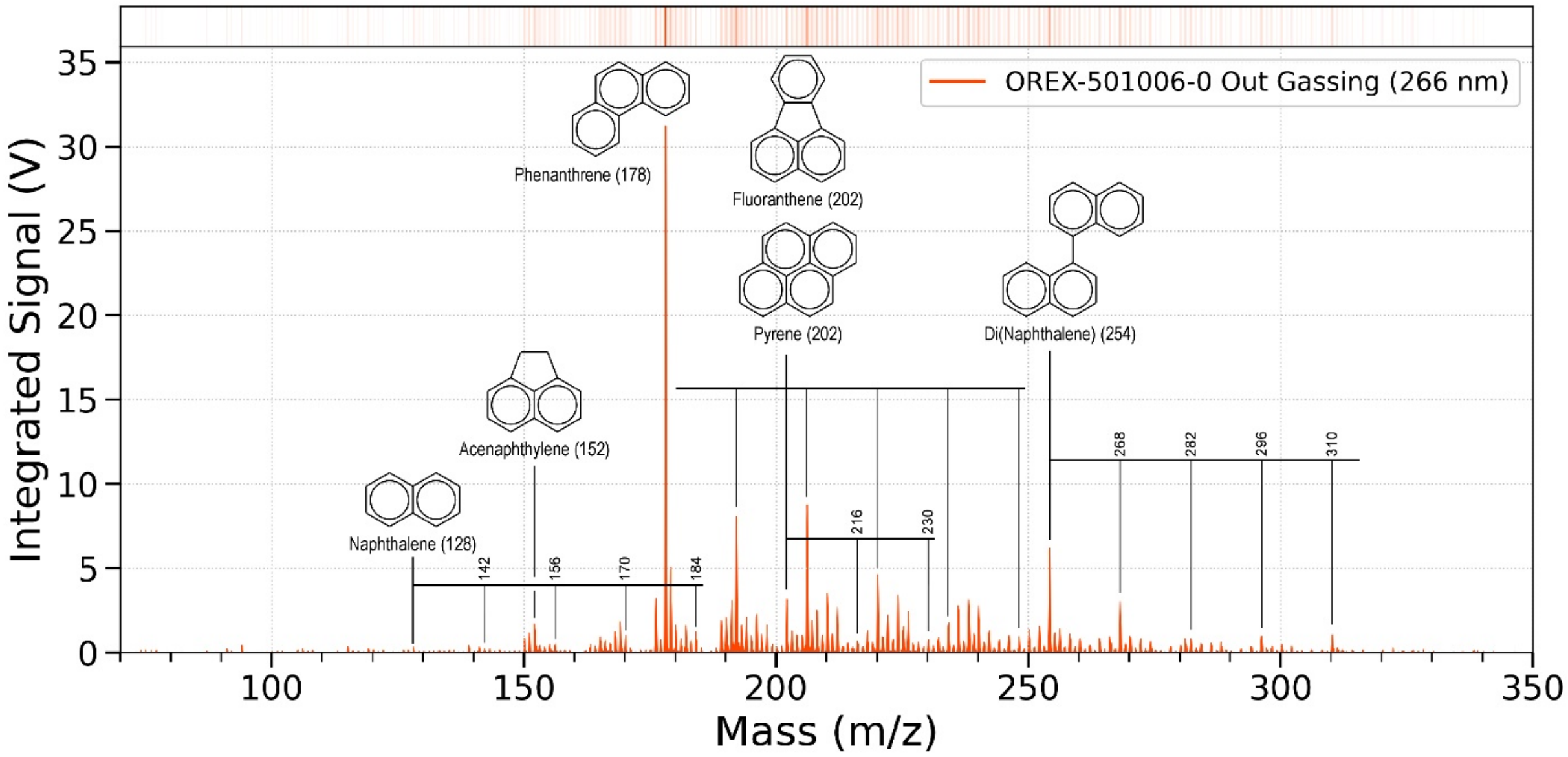

**Figure 21: µL2MS spectrum of organic molecules that outgassed from the sample after loading into the vacuum chamber**. The spectrum of the released molecules after laser desorption is identical.

### *Presolar Grains*

To search for presolar grains, nanoscale secondary ion mass spectrometry (NanoSIMS) analyses of a QL sample (Table 1) were performed using the CAMECA NanoSIMS 50L at JSC. Isotopes of C, N, and Si were simultaneously mapped over small regions of the sample at approximately 100 nm spatial resolution.  Presolar silicon carbide (SiC) grains and presolar graphite were identified. Based on the observed distribution of the presolar grains, we determined abundances of 52 (−10/+12) for presolar SiC and 12 (−5/+7) ppm for presolar graphite.

The compositions of the presolar SiC grains, in comparison to literature data (Zinner 2003), indicates that most of them are mainstream grains consistent with an origin in low-mass asymptotic giant branch (AGB) stars with solar metallicity (Figure 22). One grain enriched in both $^{13}C$ and $^{15}N$ falls into the X classification, signifying its origin from a supernova. Also notable is one presolar SiC, about 300 nm in size, that is remarkably enriched in $^{13}C$, surpassing the solar composition by approximately 17.5 times. Classified as a type-AB grain, its potential origins span J-type C stars, born-again AGB stars, and type-II supernovae.

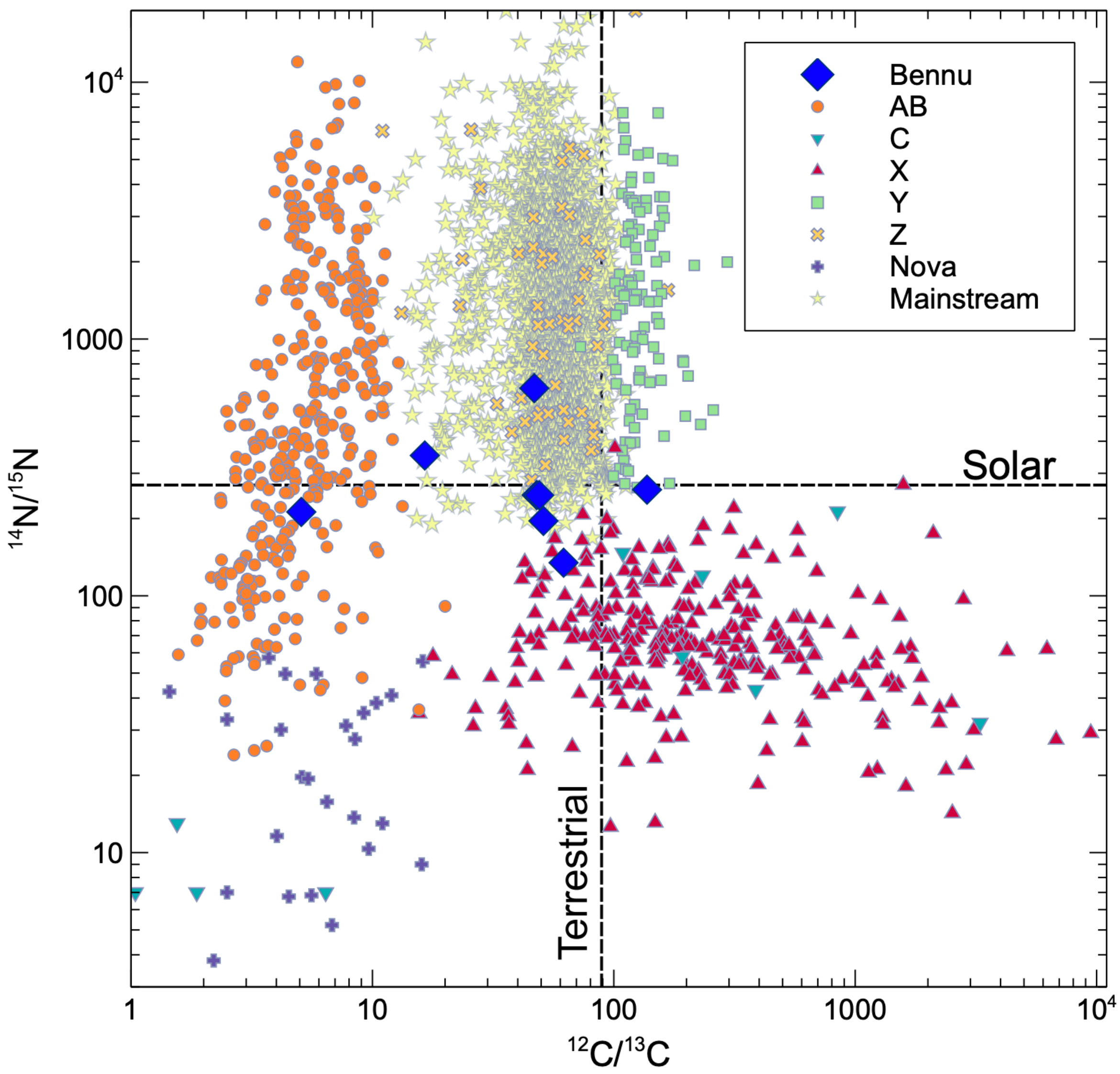

**Figure 22: Isotopic compositions of presolar grains identified in a QL sample of Bennu.** The isotopic compositions of known types of presolar grains are plotted for comparison.

## Discussion

### Testing Hypothesis 1: Comparison to Spacecraft Remote Sensing Data and Carbonaceous Astromaterials

The sample analysis campaign for OSIRIS-REx follows a hypothesis-driven approach, using findings from the asteroid encounter to structure investigations into Bennu's composition and history, spanning from pre-accretion to the present (Lauretta et al., 2023b). The primary hypothesis addressed in this manuscript, designated as Hypothesis 1 in Lauretta et al. (2023b), posits that remote sensing has effectively characterized Bennu's surface mineralogy, chemistry, and physical properties. This overarching hypothesis is further divided into five sub-hypotheses.

#### *Hypothesis 1.1: Two primary lithologies*

Hypothesis 1.1 proposes that Bennu's surface composition can be categorized into two primary lithologies based on their albedo, thermal and physical properties, and spectral slope (DellaGiustina et al., 2020, Rozitis et al., 2020). We address this hypothesis using the analyses of particle morphology, density, and qualitative reflectance.

Among the low-reflectance particles, the presence of two main morphologies — hummocky and angular — with distinct densities (Table 2) supports the existence of at least two lithologies. The hummocky stones likely correspond to the dark type A boulders on Bennu (Jawin et al., 2023), while the angular stones have smooth faces that resemble the brighter type C boulders (Jawin et al., 2023). It is worth nothing that type A and C boulders belong to populations with similar but quantitatively distinct reflectance, whereas the hummocky and angular stones do not differ meaningfully in reflectance as observed by eye — though this could be because the differences are too subtle for visual detection. Quantitative measurements of reflectance and thermal inertia will provide insights into potential correlations with boulders and explore the apparently fractal nature observed, where particle shapes at smaller scales mimic boulders at larger scales.

#### *Hypothesis 1.2: Consistency of bulk mineralogy and chemistry with remote sensing interpretations*

Hypothesis 1.2 anticipates the presence of specific minerals and compounds in the sample, including Mg-rich phyllosilicates, magnetite, complex organic mixtures, carbonates, and <10% anhydrous silicates. Sulfide minerals also inferred based on analog meteorites (Lauretta et al., 2023b; Hamilton et al., 2021).

Our investigations confirm the dominant presence of abundant Mg-rich phyllosilicates, primarily composed of serpentine- and smectite-group minerals. These phases exhibit diverse morphologies from platy to wispy structures. Sulfides, predominantly iron and sulfur with minor nickel, are widely distributed among the aggregate samples studied, at sizes from sub-micron to several microns in diameter. Many of the sulfides exhibit euhedral hexagonal plaquette shapes, consistent with pyrrhotite. Magnetite is present in various morphologies, including framboids, rosettes, and plaquettes, all exhibiting surface pitting indicative of late-stage corrosion.

Carbon is present in both minerals and organic molecules. Carbonates containing calcium, magnesium, and manganese are identified, corresponding to calcite, dolomite, and breunnerite. Micron-sized carbon-rich domains are dispersed throughout the matrix as nanoglobules, along with organic materials in plate, vein, and aggregate forms.

Additional minor phases include olivine, pyroxene, and spinel (chromite and ilmenite). The relatively low abundance of these anhydrous silicates and oxides is consistent with the prediction of <10% of such material from the remote observations (Hamilton et al., 2021).

Thus, the sample's mineralogical diversity supports Hypothesis 1.2. Our analyses demonstrate the accuracy of spacecraft remote sensing instruments in assessing Bennu's surface mineralogy.

However, the presence of phosphate rather than carbonate as the bright material in the mottled stone analyzed here is unexpected, given the spectral evidence of carbonate-rich bright veins and the lack of phosphate detection on Bennu (Kaplan et al., 2020). It is plausible that the phosphate crusts and veins within the sample are components of, or related to, the bright veins found in Bennu's boulders. The mottled stones show the highest measured densities (Table 2), consistent with the hypothesis that the vein-rich regions within type C and D boulders are denser than the host rock. However, further investigations are necessary to establish the abundance, distribution, and petrography of carbonates, phosphates, and other phases in the sample before conclusions can be drawn about the composition of the high-reflectance veins observed on Bennu.

### *Hypothesis 1.3: Presence of exogenic material*

Hypothesis 1.3 posits that Bennu samples may contain material originating from V-type (DellaGiustina et al., 2021) and/or S-type asteroids (Le Corre et al., 2021; Tatsumi et al., 2021). To date, no conclusive evidence of exogenic material has been identified in the returned samples.

### *Hypothesis 1.4: Resemblance to aqueously altered carbonaceous chondrites*

Hypothesis 1.4 proposes that the dominant lithologies on Bennu's surface have mineralogical, petrological, and compositional properties closely resembling those of the most aqueously altered carbonaceous chondrites. Here we compare Bennu samples to various aspects of the low-petrologic-type carbonaceous chondrites, as well as the Ryugu samples returned by Hayabusa2.

Our XRD analysis indicates that Bennu is very similar to the CIs and Tarda, though distinct from the CM1s (King et al., 2017). In their XRD study, King et al. (2015) determined that the CI chondrites Alais, Orgueil, and Ivuna are primarily composed of mixed serpentine-smectite phyllosilicates (81–84 vol%), with minor amounts of magnetite (6–10%), sulfides (4–7%), and carbonates (<3%). These values are similar to those found in our PXRD analysis of Bennu samples: 80% phyllosilicates, 10% sulfides, 5% magnetite, and 3% carbonates by volume.

Different carbonaceous chondrite groups exhibit distinct volatile and moderately volatile elemental compositions (Bland et al., 2005). In contrast to CI chondrites, which closely match the composition of the solar photosphere, other carbonaceous chondrite groups show varying degrees of depletion in volatile elements and corresponding enrichments in refractory elements. The extent of volatile element depletion is linked to the 50% condensation temperature (Lodders 2003). Understanding these elemental patterns is crucial for identifying chondrite groups and elucidating chondrite formation processes. Therefore, understanding the bulk chemical composition of Bennu is key to uncovering its history and placing it within the context of known astromaterials.

Compositionally, the aggregate sample from inside TAGSAM used for bulk elemental analysis closely matches the CI chondrites and Ryugu samples for the 54 elements analyzed in this study. However, the QL sample from the avionics deck shows enrichment in 10 elements compared to both the TAGSAM sample and the CI chondrites, suggesting the presence of a second chemical component in the QL sample that is distinct from solar composition. These enriched elements are well known for their fluid mobility, suggesting that their enrichment may have occurred through an open chemical system, concentrating these fluid-mobile elements during hydrothermal alteration. An open hydrothermal system is consistent with the analysis of veins in Bennu boulders by Kaplan et al. (2020).

The Bennu samples we analyzed are enriched in bulk C compared to carbonaceous chondrites and the average of samples returned from Ryugu. The bulk isotopic compositions of H and N in Bennu samples show slight distinctions from carbonaceous chondrites and Ryugu samples. Though falling within the range of carbonaceous astromaterials, Bennu's δ15N and δD compositions align more closely with those of the ungrouped carbonaceous meteorites Tarda and Tagish Lake than with other meteorites.

The average oxygen isotopic composition that we measured places Bennu in the same region of oxygen three-isotope space as some of the most chemically primitive chondritic materials known (Clayton and Mayeda 1999), including CI and CY chondrites, as well as samples from Ryugu. However, the average $\Delta^{17}O$ value is slightly higher than that of other meteorite groups or returned samples.

The abundances of presolar SiC and graphite grains align closely with those found in unheated chondrite samples (Huss and Lewis 1995). These discoveries are consistent with previous findings from Ryugu, indicating a possible shared cosmic heritage between these asteroids (Nguyen et al., 2023).

In summary, the comparison of Bennu samples to carbonaceous chondrites and Ryugu materials provides compelling evidence supporting Hypothesis 1.4. Although significant parallels exist between Bennu and carbonaceous astromaterials, subtle variations in the isotopic ratios and our observation of a distinct chemical component enriched in fluid-mobile elements underscore the unique nature of these samples and the complex processes shaping asteroidal evolution still to be uncovered through extended sample analysis.

### *Hypothesis 1.5: Physical properties distinct from meteorites*

The physical properties of the Bennu samples were anticipated to deviate from those observed in meteorites. This divergence is exemplified by the generally low thermal inertia of boulders, particularly in the “dark” population (DellaGiustina et al., 2020), which is inferred to indicate high-porosity, low-density material that probably would not survive Earth atmospheric entry (Rozitis et al., 2020). The bulk densities of stones from carbonaceous chondrites vary widely (Macke et al., 2011) — as low as 1.57 ± 0.03 g/cm$^3$ for the CI Orgueil and as high as 5.55 ± 0.09 g/cm$^3$ for Bencubbin, a metal-rich CB chondrite. Some of the bulk densities of hummocky Bennu particles measured in the laboratory are like those of Orgueil (Table 2). This result suggests that Bennu's regolith particles have a similar density to known carbonaceous chondrites.

The average bulk density of 637 Ryugu particles was determined to be 1.79 ± 0.31 g/cm$^3$ for weights ranging from 0.5 to 100 mg (sub-millimeter to 10 mm), which is considered representative of the returned samples (Miyazaki et al., 2023). The bulk density distributions of the particles in the two different Hayabusa2 collection chambers were statistically distinguishable, with mean values of 1.81 ± 0.30 and 1.76 ± 0.33 g/cm$^3$, respectively. These values are higher than the average density of the angular particles we measured, but like that of the mottled particles (Table 2).

The extensive petrographic similarity between the Bennu and Ryugu samples, both composed of material resembling the rarest of meteorite classes, suggests that low-density, friable material is abundant in near-Earth space. However, this material is unlikely to survive passage through Earth’s atmosphere (Baldwin et al., 1971; Rozitis et al., 2020). This highlights the value of sample return missions in providing a comprehensive understanding of solar system materials.

## Insights Into the Formation of the Bennu Samples

We have identified serpentinites as the dominant rock type. Comparing these serpentinites with their terrestrial counterparts can provide insights into Bennu's geological past. On Earth, serpentinites form when ultramafic rocks undergo hydration and alteration, processes influenced by factors such as temperature, pressure, water availability, and the presence of $CO_2$-rich fluids (Deschamps et al. 2013; Evans et al. 2013). Understanding these conditions during serpentinization on Bennu's parent asteroid is crucial for deciphering its geological evolution. The morphologies of the particles may offer clues about the aqueous environment in which they originated.

On Earth, low-density, hummocky rocks are typically formed through the deposition and erosion of loose sediments. A geological example can be seen in Guardavalle, Calabria, southern Italy, where two Pliocene sequences exhibit distinctive intermediate-scale lamination with hummocky cross-stratification (DeCelles et al., 1992). These sedimentary features indicate a shallow depositional environment, with depths ranging from 2 to 5 m.

The similarities between the rocks in Calabria and Bennu's type A boulders and hummocky particles suggest common depositional processes. In both cases, the rocks display hummocky textures, cross-stratification, linear layer boundaries, and parallel bedding (DeCelles et al., 1992; Ishimaru and Lauretta, 2023). These findings suggest that the hummocky textures on Bennu may have formed through settling from suspension, likely due to unsteady flows. This scenario aligns with the proposed sedimentary deposition as the origin of layered boulders on Bennu (Ishimaru and Lauretta, 2023).

In contrast, the higher-density, angular particles from Bennu resemble shapes found in terrestrial serpentinite quarries during rock fracturing (Marescotti et al., 2006). These angular shapes can be attributed to columnar serpentine textures, which form when mineral growth is constrained, leading to elongated, column-like shapes (Boudier et al., 2010). Specifically, columnar lizardite forms when hydrous fluids penetrate peridotite through initial cracks, triggering nucleation and growth from the crack towards the center of the host olivine grains. As serpentinization progresses inward with increasing water/rock ratio, the columnar morphology is maintained. Therefore, the morphology of angular Bennu particles may indicate serpentinization in a volumetrically constrained, possibly higher-pressure environment compared to the hummocky particles.

The discovery of magnesium-sodium phosphate material in the Bennu samples provides insights into fluid chemistry. This material, resembling sodium phosphates found in Saturn's moon Enceladus (Postberg et al., 2023), suggests a possible link between these bodies. Enceladus's sodium-rich phosphates, discovered in its global ice-covered water ocean and cryovolcanic plumes, indicate high pH values, like those predicted for fluids responsible for altering carbonaceous chondrites (Zolensky et al., 1989). Similar phosphate-enrichment processes in basic fluids are observed in Earth's soda lakes, such as Last Chance Lake and Goodenough Lake in Canada, where high concentrations of dissolved phosphate result from the precipitation of calcium-rich carbonate (Haas et al., 2024). These findings highlight the complex geochemical processes leading to phosphate enrichment in aqueous systems, emphasizing the importance of carbonate minerals and the solubility of phosphate-rich phases in facilitating phosphate accumulation. These studies suggest that the texture observed in the mottled particles may have formed when cracks and voids were filled with phosphates during post-serpentinization metasomatism followed by evaporite formation. This event would have changed the bulk composition of the material, introducing elements that are mobile in hydrothermal environments. Detailed studies of the carbonates, phosphates, and potential other vein-filling materials in the Bennu samples are needed to understand the complex fluid chemistry recorded in these stones.

## Conclusions

Our initial analyses of the Bennu samples largely confirm predictions from telescopic and spacecraft data. The sample is hydrated (0.84–0.95 wt.% H) and carbon-rich (4.5–4.7 wt.% C), with bulk elemental abundances similar to solar composition. Its mineralogy is like that of primitive, aqueously altered carbonaceous astromaterials, comprising (in approximately decreasing order of abundance): phyllosilicates (serpentine and smectite-group minerals); Fe,Ni-sulfides, including pyrrhotite and pentlandite; magnetite; carbonates such as calcite, dolomite, and breunnerite; forsteritic olivine; low-Ca pyroxene; Mg,Na- and Ca-phosphates; and other trace phases. The organic nanoglobules and PAHs that we identified do not show evidence of consequential thermal processing. Presolar carbides and graphite are preserved. Together, these findings suggest that the materials composing Bennu's regolith, though extensively aqueously altered, have retained some signatures from the protoplanetary disk.

The hydrated Mg,Na-phosphates, with similarities to sodium phosphates on Enceladus and in Earth's soda lakes, suggest that complex fluid chemistry may have been at play. This fluid could have introduced a distinct chemical component, leading to the deviation from solar composition that we observed in the bulk elemental abundances of one subsample. It is noteworthy that phosphates were not detected in the spacecraft data or in every sample, which may have implications for our understanding of Bennu's mineralogical composition and formation history.

The distinct hydrogen, carbon, and nitrogen isotopic compositions of Bennu's regolith raise questions about its formation and evolution. Further investigation into these isotopic signatures could provide valuable insights into the history of Bennu and its parent body.

The three predominant types of stones in the sample — hummocky, angular, and mottled — corroborate the existence of different boulder types on Bennu, particularly given their distinct densities. The hummocky and angular stones resemble boulder morphologies observed in spacecraft images, and the higher reflectance of the mottled stones suggests a potential link to the bright veins evident in some boulders. However, analyses of quantitative reflectance, thermal properties, composition, and mineralogy are needed to substantiate these connections.

Our first-look findings highlight the importance of sample return missions in unraveling the geological and geochemical intricacies of asteroids like Bennu — whose low-density materials are probably underrepresented in the meteorite record — and their implications for the formation and evolution of the solar system. Nonetheless, the data we have presented here are only the tip of the iceberg: there is likely more about the sample that we do not know than we do know.

The OSIRIS-REx team is actively testing the driving hypotheses that guide our analytical approach (Lauretta et al., 2023b), exploring in detail the sample's petrology and petrography, mineralogical and organic components, elemental and isotopic compositions, thermal and physical properties, and exposure to space weathering to build a comprehensive geologic history of Bennu's regolith. Our findings will provide the foundation for further analysis by the broader community via the publicly available Bennu sample catalog.

## Acknowledgments

This material is based upon work supported by NASA under Award NNH09ZDA007O and Contract NNM10AA11C issued through the New Frontiers Program. We are grateful to the past and present membership of the OSIRIS-REx Team who made the return of samples from Bennu possible. We thank the Astromaterials Acquisition and Curation Office, part of the Astromaterials Research and Exploration Science (ARES) Division at Johnson Space Center, for their efforts in SRC recovery, preliminary examination, and long-term curation. MP and FT were supported for this research by the Italian Space Agency (ASI) under the ASI-INAF agreement no. 2022-1-HH.0. SSR and AJK acknowledge support from the Science and Technology Facilities Council (STFC) of the UK. IAF and RCG thank STFC for funding (grant ST/Y000188/1). ST is supported by JSPS KAKENHI Grant Number 20H05846 and 22K21344.

## Data Availability

The OSIRIS-REx sample analysis data that support the findings of this study will be available via AstroMat at the DOIs given in the table below.

**AIVA Images**

| DOI | Product Name | Product Type |
|---|---|---|
| **AIVA** | Canister, TAGSAM, Sample in Trays | |
| **10.60707/ek2k-3171** | 20231020_AIVA_JSC-ARES_multiSample_1_AIVAImage_1.tif | AIVAImage |
| **10.60707/7fk7-4x67** | 20240122_AIVA_JSC-ARES_multiSample_1_AIVAImage_1.tif | AIVAImage |
| **10.60707/k0dp-cp12** | 20230926_AIVA_JSC-ARES_OREX-108001-0_1_AIVAImage_1.tif | AIVAImage |

**EA-IRMS Data**

| DOI | Product Name | Product Type |
|---|---|---|
| **EA-IRMS** | OREX-501033-0, OREX-501034-0, OREX-501035-0, OREX-501036-0, OREX-501037-0, OREX-501038-0, OREX-501039-0, OREX-501040-0, OREX-501041-0 | |
| **10.60707/wg35-6e70** | 20231005_EA-IRMS_CIS_OREX-501033-0_1_EAIRMSCollection_1.zip | EAIRMSCollection |
| **10.60707/6c5n-e486** | 20231005_EA-IRMS_CIS_multiSample_1_EAIRMSCollection_1.zip | EAIRMSCollection |
| **10.60707/t5ac-es57** | 20231004_EA-IRMS_CIS_multiSample_1_EAIRMSCollection_1.zip | EAIRMSCollection |
| | OREX-803040-0, OREX-803041-0, OREX-803042-0, OREX-803043-0, OREX-803002-0, OREX-803044-0, OREX-803045-0, OREX-803046-0 | |
| **10.60707/ndf3-qn80** | 20231210_EA-IRMS_CIS_multiSample_1_EAIRMSCollection_1.zip | EAIRMSCollection |
| **10.60707/0g1m-4v39** | 20231209_EA-IRMS_CIS_multiSample_1_EAIRMSCollection_1.zip | EAIRMSCollection |

**XCT Data**

| DOI | Product Name | Product Type |
|---|---|---|
| **XCT** | OREX-500003-100 | |
| **10.60707/7k9y-n295** | 20231002_XCT_JSC-ARES_OREX-500003-100_1_XCTImageCollection_1.zip | XCTImageCollection |

**SEM/EDS Data**

| DOI | Product Name | Product Type |
|---|---|---|
| **SEM/EDS** | OREX-501001-0 | |
| | OREX-501002-0 | |
| | OREX-501017-0 | |
| | OREX-803079-0 | |
| | OREX-803080-0 | |
| | OREX-803100-0 | |
| | | |
| | OREX-803009-101 | |
| **10.60707/hbmk-tz71** | 20231205_SEM_UAZ_OREX-803009-101_1_SEMImageMap_3.tif | SEMImageMap |
| **10.60707/srdj-pr74** | 20231205_SEM_UAZ_OREX-803009-101_1_SEMImageMap_2.tif | SEMImageMap |
| **10.60707/9z95-a638** | 20231205_SEM_UAZ_OREX-803009-101_1_SEMImageMap_1.tif | SEMImageMap |
| **10.60707/9cyc-q314** | 20231204_VLM_UAZ_OREX-803009-101_1_VLMImage_1.tif | VLMImage |

**UV-L2MS-μFTIR Data**

| DOI | Product Name | Product Type |
|---|---|---|
| **UV-L2MS-μFTIR** | OREX-501006-0 | |
| | OREX-501018-0 | |
| | OREX-501000-0 | |

**XRD Data**

| DOI | Product Name | Product Type |
|---|---|---|
| **XRD** | OREX-500005-0 | |
| **10.60707/2j0t-gq80** | 20230928_XRD_JSC-ARES_OREX-500005-0_1_XRDTabular_1.csv | XRDTabular |

**ICP-MS Data**

| DOI | Product Name | Product Type |
|---|---|---|
| **ICP-MS** | OREX-501043-0 | |
| **10.60707/w6ea-6e34** | 20240129_MC-ICP-MS_WUSTL_multiSample_1_MCICPMSCollection_1.zip | MCICPMSCollection |
| **10.60707/xd7z-kv85** | 20240129_MC-ICP-MS_WUSTL_multiSample_1_MCICPMSTabular_1.csv | MCICPMSTabular |
| **10.60707/kw6s-c650** | 20231128_Q-ICP-MS_WUSTL_multiSample_1_QICPMSProcessedTabular_1.csv | QICPMSProcessedTabular |
| **10.60707/s2qg-6q60** | 20231128_Q-ICP-MS_WUSTL_multiSample_1_QICPMSRawTabular_1.csv | QICPMSRawTabular |
| **10.60707/xyvj-4s46** | 20231103_Q-ICP-MS_WUSTL_OREX-501043-0_1_QICPMSProcessedTabular_1.csv | QICPMSProcessedTabular |
| **10.60707/n62d-z186** | 20231103_Q-ICP-MS_WUSTL_OREX-501043-0_1_QICPMSRawTabular_1.csv | QICPMSRawTabular |
| **10.60707/gkfw-1m44** | 20231015_Q-ICP-MS_WUSTL_OREX-501043-0_1_QICPMSProcessedTabular_1.csv | QICPMSProcessedTabular |
| **10.60707/17dy-df56** | 20231015_Q-ICP-MS_WUSTL_OREX-501043-0_1_QICPMSRawTabular_1.csv | QICPMSRawTabular |
| | OREX-803015-0 | |
| **10.60707/w6ea-6e34** | 20240129_MC-ICP-MS_WUSTL_multiSample_1_MCICPMSCollection_1.zip | MCICPMSCollection |
| **10.60707/xd7z-kv85** | 20240129_MC-ICP-MS_WUSTL_multiSample_1_MCICPMSTabular_1.csv | MCICPMSTabular |
| **10.60707/kw6s-c650** | 20231128_Q-ICP-MS_WUSTL_multiSample_1_QICPMSProcessedTabular_1.csv | QICPMSProcessedTabular |
| **10.60707/s2qg-6q60** | 20231128_Q-ICP-MS_WUSTL_multiSample_1_QICPMSRawTabular_1.csv | QICPMSRawTabular |
| **10.60707/t3a6-mt98** | 20231128_Q-ICP-MS_WUSTL_OREX-803015-0_1_QICPMSProcessedTabular_1.csv | QICPMSProcessedTabular |
| **10.60707/4kta-vg28** | 20231128_Q-ICP-MS_WUSTL_OREX-803015-0_1_QICPMSRawTabular_1.csv | QICPMSRawTabular |

**LF O-isotope Data**

| DOI | Product Name | Product Type |
|---|---|---|
| **LF** | OREX-501042-0 | |
| **10.60707/7tn1-6359** | 20240124_LAF_OU_OREX-501042-0_1_LAFProcessed_1.CSV | LAFProcessed |
| **10.60707/2rbs-7m49** | 20240124_LAF_OU_OREX-501042-0_1_LAFRaw_1.zip | LAFRaw |
| | OREX-501047-0 | |
| **10.60707/km0m-0p38** | 20240124_LAF_OU_OREX-501047-0_1_LAFProcessed_1.CSV | LAFProcessed |
| **10.60707/th82-qz83** | 20240124_LAF_OU_OREX-501047-0_1_LAFRaw_1.zip | LAFRaw |
| | OREX-501066-0 | |
| **10.60707/w65d-v386** | 20240123_LAF_OU_OREX-501066-0_1_LAFProcessed_1.CSV | LAFProcessed |
| **10.60707/z96t-9p57** | 20240123_LAF_OU_OREX-501066-0_1_LAFRaw_1.zip | LAFRaw |
| | OREX-501067-0 | |
| **10.60707/jmnb-rn32** | 20240124_LAF_OU_OREX-501067-0_1_LAFProcessed_1.CSV | LAFProcessed |
| **10.60707/k39c-sx93** | 20240124_LAF_OU_OREX-501067-0_1_LAFRaw_1.zip | LAFRaw |

**NanoSIMS Data**

| DOI | Product Name | Product Type |
|---|---|---|
| **NanoSIMS** | OREX-501018-100 | |
| **10.60707/nm1b-bz91** | 20231221_NanoSIMS_JSC-ARES_OREX-501018-100_1_NanoSIMSImageCollection_1.zip | NanoSIMSImageCollection |
| **10.60707/jj7q-6v48** | 20231121_NanoSIMS_JSC-ARES_OREX-501018-100_1_NanoSIMSImageCollection_1.zip | NanoSIMSImageCollection |
| **10.60707/zz81-fv27** | 20231114_NanoSIMS_JSC-ARES_OREX-501018-100_1_NanoSIMSImageCollection_1.zip | NanoSIMSImageCollection |

**RELAB Spectral Data**

| DOI | Product Name | Product Type |
|---|---|---|
| **RELAB** | OREX-800029-0 | |
| **10.60707/n69q-jm35** | 20231113_VNMIR_BROWN_OREX-800029-0_1_VNMIRSpectralPoint_1.csv | VNMIRSpectralPoint |
| **10.60707/wset-j646** | 20231113_VNMIR_BROWN_OREX-800029-0_1_VNMIRSpectralPoint_2.csv | VNMIRSpectralPoint |
| **10.60707/dqts-2513** | 20231111_VNMIR_BROWN_OREX-800029-0_1_VNMIRSpectralPoint_1.csv | VNMIRSpectralPoint |
| **10.60707/8xja-1723** | 20231111_VNMIR_BROWN_OREX-800029-0_1_VNMIRSpectralPoint_2.csv | VNMIRSpectralPoint |
| **10.60707/pw1a-2296** | 20231110_VNMIR_BROWN_OREX-800029-0_1_VNMIRSpectralPoint_1.csv | VNMIRSpectralPoint |

**SLS Data**

| DOI | Product Name | Product Type |
|---|---|---|
| **SLS** | OREX-800014-0 | |
| | OREX-800017-0 | |
| | OREX-800019-0 | |
| | OREX-800020-0 | |
| | OREX-800021-0 | |
| | OREX-800023-0 | |
| | OREX-800026-0 | |
| | OREX-800055-0 | |
| | OREX-800067-0 | |
| | OREX-800087-0 | |
| | OREX-800088-0 | |
| | OREX-800089-0 | |

The image mosaics of Bennu and Nightingale in Figure 1 are available here:
https://www.asteroidmission.org/3_asteroid-bennu-rotation_sm/
https://www.asteroidmission.org/nightingale-recon-a-mosaic/
https://www.asteroidmission.org/candidate-sample-sites/nightingale/nightingalerecona/#main

The following images used in Figure 2 are available from the NASA Planetary Data System, Small Bodies Node.

| **PolyCam Images** |
|---|
| 20190328T182843S131_pol_iofL2pan_V005 |
| 20190405T203448S024_pol_iofL2pan_V007 |
| 20190412T174942S616_pol_iofL2pan_V003 |
| 20190412T180648S629_pol_iofL2pan_V003 |
| 20190412T184031S353_pol_iofL2pan_V003 |
| 20191026T211251S139_pol_iofL2pan_V011 |
| 20191026T213446S707_pol_iofL2pan_V012 |

20191026T214514S803_pol_iofL2pan_V012

| Name | Institution | Institution Country |
| --- | --- | --- |
| Alasli, Abdulkareem | Nagoya University | Japan |
| Alexander, Conel M. O'D. | Carnegie Institution for Science | USA |
| Alexandre, Marcelo | Brown University | USA |
| Allums, Kimberly | NASA-JSC | USA |
| Almeida, Natasha | Natural History Museum | United Kingdom |
| Amini, Marghaleray | University of British Columbia | Canada |
| Aponte, Jose | NASA-GSFC | USA |
| Applin, Daniel | University of Winnipeg | Canada |
| Asphaug, Erik | University of Arizona | USA |
| Avdellidou, Chrysa | Observatoire de la Côte d'Azur/CNRS | France |
| Azéma, Emilien | Univ. Montpellier | France |
| Baczynski, Allison A | Pennsylvania State University | USA |
| Bajo, Ken-ichi | Hokkaido University | Japan |
| Ballouz, Ronald | Johns Hopkins University Applied Physics Laboratory | USA |
| Barnes, Jessica | University of Arizona | USA |
| Barnouin, Olivier | Johns Hopkins University Applied Physics Laboratory | USA |
| Bates, Helena | Natural History Museum | United Kingdom |
| Bechtel, Hans | Lawrence Berkeley National Laboratory | USA |
| Bekaert, David | Centre de Recherches Pétrographiques et Géochimiques | France |
| Bennett, Carina | University of Arizona | USA |
| Berger, Eve | NASA-JSC | USA |
| Berkson, Michael | Johns Hopkins University Applied Physics Laboratory | USA |
| Biele, Jens | German Aerospace Center (DLR) | Germany |
| Bierhaus, Beau | Lockheed Martin Corporation | USA |
| Blanche, Laurinne | University of Arizona | USA |
| Bland, Phil | Curtin University | Australia |
| Blum, Denise | University of Arizona | USA |
| Borg, Lars Eric | Lawrence Livermore National Laboratory/Dept of Energy | USA |
| Bowles, Neil | University of Oxford | United Kingdom |
| Boynton, William | University of Arizona | USA |
| Brenker, Frank | Goethe University Frankfurt | Germany |
| Brennecka, Greg | Lawrence Livermore National Laboratory/Dept of Energy | USA |
| Burgess, Raymond | Manchester University | United Kingdom |
| Burton, Aaron | NASA-JSC | USA |
| Busemann, Henner | ETH Zürich | Switzerland |
| Caffee, Marc W. | Purdue University | USA |
| Chaves, Laura | University of Arizona | USA |
| Chikaraishi, Yoshito | Hokkaido University | Japan |
| Chung, Angela | Catholic University of America | USA |
| Calva, Curtis | Jacobs/Jets (NASA JSC) | USA |
| Ciceri, Fabio | University of Calgary | Canada |
| Clark, Ben | Space Science | USA |
| Clay, Patricia | University of Ottawa | Canada |
| Clemett, Simon | NASA-JSC | USA |
| Cloutis, Edward | University of Winnipeg | Canada |
| Cody, George | Carnegie Institution for Science | USA |
| Connelly, Wayland | Jacobs/Jets (NASA JSC) | USA |
| Connolly, Harold | Rowan University | USA |
| Corrigan, Cari | Smithsonian Institution | USA |
| Crombie, Kate | Indigo Information Services, LLC | USA |
| Croom, Brendan | Johns Hopkins University Applied Physics Laboratory | USA |
| Crowther, Sarah | Manchester University | United Kingdom |
| Daly, Mike | York University | Canada |
| Delbo, Marco | Observatoire de la Côte d'Azur/CNRS | France |
| DellaGiustina, Daniella | University of Arizona | USA |
| Dominguez, Gerardo | University Auxiliary and Research Services Corporation | USA |
| Dukes, Cathy | University of Virginia | USA |
| Dworkin, Jason | NASA-GSFC | USA |
| Eckley, Scott | NASA-JSC | USA |
| Elsila Cook, Jamie | NASA-GSFC | USA |
| Emery, Joshua | Northern Arizona University | USA |
| Farnsworth, Kendra | University of Maryland, Baltimore County | USA |
| Feilin, Cheng | Nagoya University | Japan |
| Ferdous, Jannatul | NASA-JSC | USA |
| Ferro, Tony | University of Arizona | USA |
| Fitzgibbon, Michael | University of Arizona | USA |
| Foka, Sostehene | University of Arizona | USA |
| Foustoukos, Dionysis | Carnegie Institution for Science | USA |
| Franchi, Ian | Open University | United Kingdom |
| Francis, Scott | Lockheed Martin Corporation | USA |
| Freeman, Katherine H | Pennsylvania State University | USA |
| Freemantle, James | York University | Canada |
| Freund, Sandy | Lockheed Martin Corporation | USA |
| Fujita, Ryohei | Nagoya University | Japan |
| Fulford, Ruby | University of Arizona | USA |
| Funk, Rachel | Jacobs/Jets (NASA JSC) | USA |
| Füri, Evelyn | Centre de Recherches Pétrographiques et Géochimiques | France |
| Furukawa, Yoshihiro | Tohoku University | Japan |
| Gainsforth, Zack | University of California, Berkeley | USA |
| Gienger, Eddie | Johns Hopkins University Applied Physics Laboratory | USA |
| Gilmour, Jamie | Manchester University | United Kingdom |
| Glavin, Daniel P | NASA-GSFC | USA |
| Golish, Dathon | University of Arizona | USA |
| Gonzalez, Carla | NASA-JSC | USA |
| Grady, Monica | Open University | United Kingdom |
| Graham, Heather | NASA-GSFC | USA |
| Greenwood, Richard | Open University | United Kingdom |
| Haenecour, Pierre | University of Arizona | USA |
| Haltigin, Tim | Canadian Space Agency | Canada |
| Hamilton, Vicky | Southwest Research Institute | USA |
| Hammond, Damian | University of Arizona | USA |
| Hanton, Lincoln | University of Calgary | Canada |
| Harrington, Roger | NASA-JSC | USA |
| Heck, Philipp R | Field Museum Of Natural History | USA |
| Hernandez, Neftali | Jacobs/Jets (NASA JSC) | USA |
| Hildebrand, Alan | University of Calgary | Canada |
| Hill, Dolores | University of Arizona | USA |
| Hill, Patrick | Canadian Space Agency | Canada |
| Hiroi, Takahiro | Brown University | USA |
| Hofmann, Amy E | Jet Propulsion Laboratory | USA |
| Hoover, Christian | Arizona State University | USA |
| House, Christopher H | Pennsylvania State University | USA |
| Huang, Yongsong | Brown University | USA |
| Huss, Gary | University of Hawai'i | USA |
| Ireland, Trevor | Australian National University | Australia |
| Ishimaru, Kana | University of Arizona | USA |
| Ishizaki, Takuya | ISAS/JAXA | Japan |
| Jardine, Keanna | ASU | USA |
| Jawin, Erica | Smithsonian Institution | USA |
| Jilly-Rehak, Christine E | Stanford University | USA |
| Johnson, Mark | Lockheed Martin Corporation | USA |
| Jones, Rhian | Manchester University | United Kingdom |
| Jourdan, Fred | Curtin University | Australia |
| Kantarges, Joshua | University of Arizona | USA |
| Kaplan, Hannah | NASA-GSFC | USA |
| Kawasaki, Noriyuki | Hokkaido University | Japan |
| Keller, Lindsay | NASA-JSC | USA |
| Kelley, Mikayla | University of Arizona | USA |
| Kim, Bumsoo | Brown University | USA |
| King, Ashley J | NASA Foreign PI Support Organization | United Kingdom |
| Koefoed, Piers | Washington University | USA |
| Koga, Toshiki | Japan Agency for Marine-Earth Science & Technology (JAMSTEC) | Japan |
| Kontogiannis, Melissa | University of Arizona | USA |
| Kruijer, Thomas | Lawrence Livermore National Security, LLC | USA |
| Lai, Vivian | University of British Columbia | Canada |
| Lauretta, Dante | University of Arizona | USA |
| Le, Loan | NASA-JSC | USA |
| Libourel, Guy | Observatoire de la Côte d'Azur | France |
| Lorentson, Charles | NASA-GSFC | USA |
| Lugo, Gabriel | Jacobs/Jets (NASA JSC) | USA |
| Lunning, Nicole | NASA-JSC | USA |
| Macke, Robert | Vatican Observatory | Italy |
| Manzoni, Claudia | LSC | United Kingdom |
| Marcus, Matthew A | Lawrence Berkeley National Laboratory | USA |
| Marrocchi, Yves | Centre de Recherches Pétrographiques et Géochimiques | France |
| Martinez, Salvador | Jacobs/Jets (NASA JSC) | USA |
| Marty, Bernard | Université de Lorraine, CRPG-CNRS | France |
| Masarik, Jozef | Comenius University, Bratislava | Slovakia |
| Matney, Mila | Pennsylvania State University | USA |
| May, Brian | LSC | United Kingdom |
| McCoy, Tim | Smithsonian Institution | USA |
| McDonough, Eva | University of Arizona | USA |
| McIntosh, Ophélie | Pennsylvania State University | USA |
| McLain, Hannah | NASA-GSFC | USA |
| Melendez, Lisette | Purdue University | USA |
| Meshoulam, Alexander | Caltech | USA |
| Michel, Patrick | Observatoire de la Côte d'Azur | France |
| Milliken, Ralph | Brown University | USA |
| Mojarro, Angel | NASA-GSFC | USA |
| Molaro, Jamie | Planetary Science Institute and NASA-JPL | USA |
| Montoya, Maritza | Jacobs/Jets (NASA JSC) | USA |
| Morisset, Caroline-Emmanuelle | Canadian Space Agency | Canada |
| Nagano, Hosei | Nagoya University | Japan |
| Nagashima, Kazu | University of Hawai'i | USA |
| Naoya, Sakamoto | University of Hokkaido | Japan |
| Naraoka, Hiroshi | Kyushu University | Japan |
| Nguyen, Ann | NASA-JSC | USA |
| Nishiizumi, Kunihiko | University of California, Berkeley | USA |
| Nuevo, Michel | NASA-Ames | USA |
| Oba, Yasuhiro | Hokkaido University | Japan |
| Ogliore, Ryan C | Washington University | USA |
| Opeil, Cyril | Boston College | USA |
| Pajola, Maurizio | INAF-OAPD | Italy |
| Parker, Eric | NASA-GSFC | USA |
| Peretyazhko, Tanya | NASA-JSC | USA |
| Piani, Laurette | Centre de Recherches Pétrographiques et Géochimiques | France |
| Plummer, Julia | Jacobs/Jets (NASA JSC) | USA |
| Polit, Anjani | University of Arizona | USA |
| Portail, Marc | Centre de Recherches sur l'HétéroEpitaxie et ses Applications (CRHEA) | France |
| Ray, Soumya | Smithsonian Institution | USA |
| Reddy, Steven | Curtin University | Australia |
| Render, Jan | Lawrence Livermore National Laboratory/Dept of Energy | USA |
| Renouf, Mathieu | Univ. Montpellier | France |
| Rickard, Will | Curtin University | Australia |
| Righter, Kevin | NASA-JSC | USA |
| Rizk, Bashar | University of Arizona | USA |
| Rodriguez, Melissa | NASA-JSC | USA |
| Russell, Sara | Natural History Museum | United Kingdom |
| Ryan, Andrew J. | University of Arizona | USA |
| Sakamoto, Naoya | Hokkaido University | Japan |
| Sakatani, Naoya | ISAS/JAXA | Japan |
| Sanchez, Paul | University of Colorado | USA |
| Sandford, Scott | NASA-Ames | USA |
| Santos, Ewerton | Brown University | USA |
| Saxey, David | Curtin University | Australia |
| Schmitt-Kopplin, Philippe | Helmholtz University | Germany |
| Schoenbaechler, Maria | ETH Zürich | Switzerland |
| Schofield, Paul F | Natural History Museum | United Kingdom |
| Schultz, Cody | Brown University | USA |
| Seguin, Frederic | NASA-GSFC | USA |
| Seifert, Laura | NASA-JSC | USA |
| Shirley, Katherine | University of Oxford | United Kingdom |
| Shollenberger, Quinn R | Lawrence Livermore National Security, LLC | USA |
| Siegler, Matthew | Planetary Science Institute | USA |
| Simkus, Danielle | NASA-GSFC | USA |
| Smith, Lucas | University of Arizona | USA |
| Snead, Christopher | NASA-JSC | USA |
| Sugita, Seiji | University of Tokyo | Japan |
| Tachibana, Shogo | University of Tokyo | Japan |
| Tait, Kim | Royal Ontario Museum | Canada |
| Takano, Yoshinori | Japan Agency for Marine-Earth Science & Technology (JAMSTEC) | Japan |
| Tanaka, Satoshi | ISAS/JAXA | Japan |
| Taricco, Carla | University of Turin | Italy |
| Thomas-Keprta, Kathie | Jacobs/Jets (NASA JSC) | USA |
| Thompson, Michelle | Purdue University | USA |
| Timms, Nick | Curtin University | Australia |
| Tripathi, Havishk | Southeastern Universities Research Association | USA |
| Tsuchiyama, Akira | University of Tokyo | Japan |
| Tu, Valerie | NASA-JSC | USA |
| Tusberti, Filippo | INAF-PADOVA | Italy |
| Usui, Tomohiro | JAXA | Japan |
| Vega Santiago, Nathalia | University of Arizona | USA |
| Villeneuve, Johan | Centre de Recherches Pétrographiques et Géochimiques | France |
| Vincze, Laszlo | Ghent University | Belgium |
| Walsh, Kevin | Southwest Research Institute | USA |
| Wang, Kun | Washington University | USA |
| Warner, Justin | Johns Hopkins University Applied Physics Laboratory | USA |
| Weis, Dominique | University of British Columbia | Canada |
| Weiss, Gabriella | University of Maryland, Baltimore County | USA |
| Welten, Kees C | University of California, Berkeley | USA |
| Wessman, Andrew | University of Arizona | USA |
| Westermann, Mathilde | University of Arizona | USA |
| Westphal, Andrew J. | University of California, Berkeley | USA |
| Wilbur, Zoe | University of Arizona | USA |
| Wimpenny, Josh | Lawrence Livermore National Laboratory/Dept of Energy | USA |
| Wolner, Cat | University of Arizona | USA |
| Worsham, Emily | Lawrence Livermore National Laboratory/Dept of Energy | USA |
| Yumoto, Koki | University of Tokyo | Japan |
| Yurimoto, Hisayoshi | Hokkaido University | Japan |
| Zega, Tom | University of Arizona | USA |
| Zeszut, Zoe | University of Arizona | USA |
| Zheng, Yuke | Field Museum Of Natural History | USA |